\documentclass[aps,twocolumn,floats,prd,nofootinbib,superscriptaddress]
{revtex4-2}

\usepackage{amsmath,amsfonts,amssymb}
\usepackage{graphicx}
\usepackage{dcolumn}
\usepackage{appendix}
\usepackage{subfigure}
\usepackage{placeins}
\usepackage[usenames,dvipsnames]{xcolor} 
\usepackage[export]{adjustbox}
\usepackage{bm}
\usepackage{hyperref}
\usepackage{adjustbox}
\usepackage{tabularx}
\hypersetup{
	colorlinks=true,
	linkcolor=red,
	filecolor=magenta,      
	urlcolor=blue,
}
\usepackage[normalem]{ulem}

\def\Muv{\rm M_{UV}}

\newcommand{\RNum}[1]{\uppercase\expandafter{\romannumeral #1\relax}}

\begin{document}

\preprint{}

\title{Constraints on the fuzzy dark matter mass window from high-redshift observables}
\author{Hovav Lazare}
\email{hovavl@post.bgu.ac.il}
\affiliation{Department of Physics, Ben-Gurion University of the Negev, Be’er Sheva 84105, Israel}

\author{Jordan Flitter}
 \email{yflitter@gmail.com}
\affiliation{Department of Physics, Ben-Gurion University of the Negev, Be’er Sheva 84105, Israel}

\author{Ely D. Kovetz}
\email{kovetz@bgu.ac.il}
\affiliation{Department of Physics, Ben-Gurion University of the Negev, Be’er Sheva 84105, Israel}

\begin{abstract}
    We use a combination of high-redshift observables to extract the strongest constraints to date on the fraction of   axion fuzzy dark matter (FDM) in the mass window  $10^{-26}\,\mathrm{eV}\!\lesssim\! m_\mathrm{FDM}\!\lesssim\!10^{-23}\,\mathrm{eV}$. These observables include ultraviolet luminosity functions (UVLFs) at redshifts $4-10$ measured by the Hubble Space Telescope, a constraint on the neutral hydrogen fraction from high-redshift quasar spectroscopy, the cosmic microwave background optical depth to reionization measurement from Planck and upper bounds on the 21cm power spectrum from HERA. In order to calculate these signals for FDM cosmology, we use the {\tt 21cmFirstCLASS} code to interface between {\tt AxiCLASS} and {\tt 21cmFAST} and consistently account for the full cosmic history from recombination to reionization. To facilitate a full Bayesian likelihood analysis, we developed a machine-learning based pipeline, which is both accurate, and enables a swift statistical inference, orders of magnitude faster than a brute force approach. We find that FDM of mass $m_\mathrm{FDM} \!= \!10^{-23} \, \mathrm{eV}$ is bound to less than 16\% of the total dark matter, where the constrains strengthen towards smaller masses, reaching down to 1\% for $m_\mathrm{FDM}\! =\! 10^{-26} \, \mathrm{eV}$, both at $95\%$ confidence level. In addition, we forecast that a future detection of
     the 21cm power spectrum with HERA will lower the upper bound at  $m_\mathrm{FDM}\! =\! 10^{-23} \, \mathrm{eV}$ to $\lesssim\!1\%$.

\end{abstract}

\maketitle

\section{Introduction}
\label{sec:intro}

\begin{figure*}
    \centering
\includegraphics[width =0.925\textwidth]{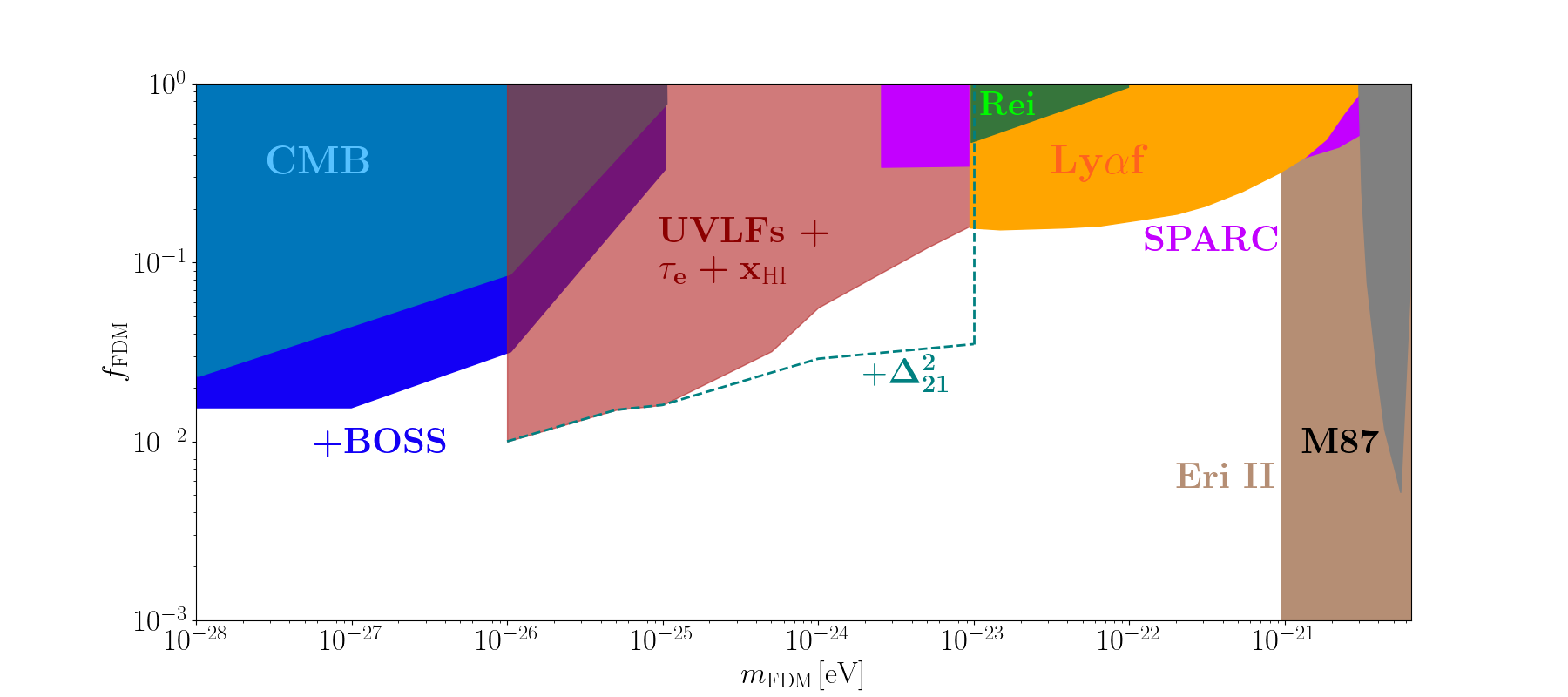}\\
\vspace{-0.05in}
\caption{Constraints on FDM  from this study, achieved using UVLFs, the CMB optical depth $\tau$, and the neutral hydrogen fraction $x_{\rm HI}$ at $z\!=\!5.9$.
For comparison, we include constraints from previous studies. In blue, CMB bounds from Planck~\cite{Hlozek:2014lca, Poulin:2018dzj}, with LSS bounds from galaxy clustering combined with Planck (+BOSS)~\cite{Lague:2021frh}. In orange, bounds from the Ly$\alpha$-forest (Ly$\alpha$f)~\cite{Irsic:2017yje, Armengaud:2017nkf, Rogers:2020ltq, Kobayashi:2017jcf}. In green, previous bounds from the UV luminosity function and optical depth to reionization (Rei.)~\cite{Bozek:2014uqa}. In purple, bounds from galaxy
rotation curves (SPARC)~\cite{Bar:2021kti}. $f_\mathrm{FDM}$ = 1 in ruled out by DES for $10^{-25}\,\mathrm{eV}\!\lesssim\! m_\mathrm{FDM}\!\lesssim\!10^{-23}\,\mathrm{eV}$ \cite{Dentler:2021zij}.
In grey, bounds from the M87 black hole spin, derived from the non-observation of superradiance~\cite{Tamburini:2019vrf, Davoudiasl:2019nlo, Unal:2020jiy}.  In brown, bounds from the half-light radius of the central star cluster in the dwarf galaxy Eridanus-II~\cite{Marsh:2018zyw}.
New constraints using UVLFs $+\,  \tau + x_{\mathrm{HI}}$ is shown in maroon, and the turquoise dashed line is a forecast for bounds that can be derived from future 21cm power spectrum observations. The last two are the main results of this work. All the bounds presented here are with 95\% confidence level.}
\vspace{-0.1in}
\label{fig:all_constraints}
\end{figure*}

While on large scales the behavior of dark matter (DM) is well understood, its small-scale characteristics remain a riddle. One of the most studied models with distinct small scale properties is that of ultra-light axions, also known as fuzzy dark matter (FDM)~\cite{Preskill:1982cy, Abbott:1982af, Dine:1982ah, Khlopov:1985jw, Chavanis:2011zi, Kawasaki:2013ae, Hlozek:2014lca, Marsh:2015xka, Marsh:2015daa,Niemeyer:2019aqm,Ferreira:2020fam,Hui:2021tkt, Hui:2016ltb, Marsh:2016vgj, Desjacques:2017fmf, Hlozek:2017zzf, Fraser:2018acy, Lidz:2018fqo, Poulin:2018dzj, Yang:2019nhz,  Bauer:2020zsj, Kawasaki:2020tbo, Schutz:2020jox, Sabiu:2021aea, Farren:2021jcd, Marsh:2021lqg, Kawasaki:2021poa, Chadha-Day:2021szb,  Boddy:2022knd, Bernal:2022jap, Laroche:2022pjm}. The particle mass landscape of FDM spans the range $10^{-27}\,\mathrm{eV}\!\lesssim\! m_\mathrm{FDM}\!\lesssim\!10^{-20}\,\mathrm{eV}$, for which the matching de-Broglie wavelengths reach Galactic scales. The quantum pressure exerted by a DM fluid made up of such tiny particles acts to suppress small scale fluctuations \cite{Marsh:2015xka, Marsh:2015daa,Niemeyer:2019aqm,Ferreira:2020fam,Hui:2021tkt}, where the smaller the mass, the stronger the suppression. 
This provides some of the theoretical appeal of this model, as it can help solve a number of issues that cast doubt on the cold DM model  when its simulations were confronted with observations, such as the core-cusp problem, the missing satellites problem and the too-big-to-fail problem \cite{Hu:2000ke, Bullock:2017xww} (although these have alternative explanations via selection effects, baryon feedback, etc.). 
Moreover, recent studies have shown that specific values of the FDM mass and fraction can potentially relax the $H_0$ \cite{Blum:2021oxj} and S$8$ \cite{Allali:2021azp,Lague:2021frh,Ye:2021iwa} tensions. FDM can also supply an alternative to the 
Stochastic Gravitational Wave Background (SGWB) explanation to the NANOGrav \cite{NANOGrav:2023gor, NANOGrav:2023hvm, NANOGrav:2023hfp} observation of pulsar timing correlations \cite{Chowdhury:2023xvy}. 
Since the FDM feature of suppressing small-scale fluctuations has a major impact on structure formation, many studies have used cosmological and astrophysical observables to constrain the FDM mass $m_{\rm FDM}$ and  fraction $f_{\rm FDM}$ parameter space. 
Such observables include cosmic microwave background (CMB) and large scale structure (LSS) measurements from Planck~\cite{Hlozek:2014lca, Lague:2021frh} and the Dark Energy Survey (DES)~\cite{DES:2018zzu}; the Ly$\alpha$-forest~\cite{Irsic:2017yje, Armengaud:2017nkf, Rogers:2020ltq, Kobayashi:2017jcf};  the high-redshift ultraviolet (UV) luminosity functions and the CMB optical depth to reionization~\cite{Bozek:2014uqa, Schive:2015kza, Corasaniti:2016epp,Winch:2024mrt,Sipple:2024svt}. 
The above limits, alongside the results of this work, are summarized in Fig.~\ref{fig:all_constraints}.
Importantly, our new constraints rule out a limited but significant mass range, $10^{-25}\,\mathrm{eV}\lesssim m_\mathrm{FDM}\lesssim10^{-23}\,\mathrm{eV}$, in which FDM was hitherto allowed to exist in significant portions. This specific region in the FDM parameter space is referred to below as the {\it FDM mass window}~\cite{Flitter:2023mjj}.

The signatures of FDM we investigate in this work are its effect on the UV luminosity functions (UVLFs), the Thomson scattering optical depth to reionization $\tau$,  the average neutral hydrogen fraction $x_{\rm HI}$ and the power spectrum of 21cm brightness temperature fluctuations from the cosmic dawn and reionization epochs. At present, the most dominant among these are the UVLFs. 

While recent studies have shown that it should be possible to close the FDM window using future 21cm measurements~\cite{Jones:2021mrs,Sarkar:2022dvl,Flitter:2023mjj} from the Hydrogen Epoch of Reionization Array (HERA)~\cite{DeBoer:2016tnn}, current HERA measurements~\cite{HERA:2021noe, HERA:2022wmy}  have only yielded upper bounds on the 21cm power spectrum in a limited range of redshifts $z$ and wavenumbers $k$, and here we find that the constraining power those upper bounds have on FDM is not competitive with other current observations. Future detection of the 21cm power spectrum, however, will produce even stronger bounds than we derive here from UVLFs+$\tau$+$x_{\rm HI}$, as we demonstrate using mock  data. 

Constraints on FDM using past measurements of the UVLFs, the CMB optical depth and the neutral  fraction were achieved to some extent in Refs.~\cite{Bozek:2014uqa, Schive:2015kza, Corasaniti:2016epp}. However, these studies used simulations and semi-analytic calculations to obtain the observables, an approach that hinders complete Bayesian statistical inference, due to the expensive computational time of the predicted signals.

Here we use updated Hubble Space Telescope (HST) observations of the UVLFs~\cite{Bouwens:2021abc} and the Planck optical depth (relative to \cite{Bozek:2014uqa, Schive:2015kza, Corasaniti:2016epp}) , which enables us to probe smaller FDM fractions compared to Refs.~\cite{Bozek:2014uqa, Schive:2015kza, Corasaniti:2016epp}. In addition, we make use of a new machine-learning (ML) based pipeline we have developed that allows us to complete a statistical inference in just a few hours.
In the traditional Bayesian markov-chain parameter-inference pipeline, realizations of the observables under examination are generated at each step, according to the likelihood of the former step.
A consistent calculation of the 21cm signal, the neutral hydrogen fraction at the EoR, and the CMB optical depth to reionization, requires simulating the evolution of the Universe from the dark ages to the EoR~\cite{Ciardi:2000wv, Santos:2009zk, Shapiro:2012xha, Hassan:2015aba, Sarkar:2016lvb, Ghara:2017vby, Mesinger:2017paq,  Sarkar:2018gcb, Ocvirk:2018pqh, Majumdar:2014cza, Sarkar:2019nak, Molaro:2019mew, Kannan:2021xoz, Shaw:2022fre, Munoz:2023kkg, Maity:2022usv, Munoz:2019fkt, Munoz:2019rhi}, which can be computationally expensive (although estimates can be achieved using fast analytical prescriptions \cite{Munoz:2023kkg}). Such simulations can take $\mathcal{O}(1\, \mathrm{hour})$ even for fast semi-numerical codes such as the public \texttt{21cmFAST} \cite{Mesinger:2010ne} and \texttt{21cmFirstCLASS}~\cite{Flitter:2023mjj} codes.
 
 Calculating the UVLFs at various redshifts can be done in $\mathcal{O}(1 \, \mathrm{sec})$ when assuming a $\Lambda {\rm CDM}$ cosmology, but when probing axion cosmology, the calculation of the matter transfer function has to be done for each set of cosmological+FDM parameters, which can take $\mathcal{O}(1 \, \mathrm{minute})$ when small scales ($k > 10\,\mathrm{Mpc}$) are considered, using a Boltzmann solver such as \texttt{AxiCLASS} \cite{Poulin:2018dzj}.  
The consequences of sequentially generating realizations for the observables we examine via this method is that a traditional Markov Chain Monte Carlo (MCMC) inference can take a significant amount of time to converge.

For that reason we have developed a machine-learning (ML) based pipeline that can overcome this obstacle.
ML techniques have emerged as powerful tools for emulating the EoR and cosmic dawn observables~\cite{Kern:2017ccn, Jennings:2018eko, Cohen:2019vck, Bye:2021ngm, Bevins:2021eah, Choudhury:2021ybn,Sikder:2022hzk, Breitman:2023pcj,Lazare:2023jkg}.
Training, validating and testing a ML model requires to pre-compute a large set of simulations, but once those are in hand
it is possible to generate fast and accurate emulators that 
given a set of cosmological and astrophysical parameters can efficiently predict the observables the model was trained on. In this work we have generated a combined emulator for all of the above-mentioned observables, and used it to evaluate the constrains they can impose on  FDM (and the FDM window in particular).

While we were finalizing a draft to summarize our results, Ref.~\cite{Winch:2024mrt} came out, where UVLFs were also used to derive bounds on FDM. In order to be able to directly compare and discuss the differences between our work and that of Ref.~\cite{Winch:2024mrt} (see Section~\ref{section_5}), we carefully revised our analysis to minimize the differences in the assumptions taken in both modeling and data selection. The consistency between our findings is encouraging given the very different analysis methods adopted. Using additional datasets, namely the neutral hydrogen  fraction and the CMB optical depth, the constraints we derive here are stronger than those reported in Ref.~\cite{Winch:2024mrt}. 

Our paper is organized as follows. In Section~\ref{section_2} we describe the computation of the different observables used in this work, and demonstrate the impact of FDM on each of them.
In Section~\ref{section_3} we describe the artificial neural network architecture our emulator is based on,  explain how the training, validation and testing sets were built, and show the emulator performance for each of the outputs. Section \ref{section_4} describes the details of our Bayesian inference scheme, presents the bounds on FDM we derived using current UVLFs, $\tau_e$ and $x_{\mathrm{HI}}$ data, and conducts a forecast for constraints that can be achieved from future 21cm power spectra measurements. In Section~\ref{section_5} we discuss our results and compare them to other works.

\section{Computing and simulating the observables}
\label{section_2}
To probe FDM, we  combine four separate observables: (i) the faint galaxy UVLFs from high redshifts, as measured by HST~\cite{Bouwens:2021abc, Gillet:2019fjd}; (ii) the Planck~\cite{Planck:2018vyg} bound on the Thomson scattering optical depth of CMB photons, $\tau_e=0.0569^{+0.0081}_{-0.0066}$, taken from Ref.~\cite{Qin:2020xrg}; (iii) the neutral hydrogen fraction  measured by the dark (zero-flux) fraction in high redshift quasar spectra, $x_{\rm HI} < 0.06 + 0.05$ (1$\sigma$) at redshift $z=5.9$~\cite{McGreer:2014qwa}; and (iv) the 21cm power spectrum measured by HERA~\cite{HERA:2022wmy}. 
In the following, we will explain how each of them is computed, given a set of  astrophysical, cosmological and FDM parameters. 

The computation is done using
 \texttt{21cmFirstCLASS}~\cite{Flitter:2023mjj, Flitter:2023rzv}, an extended version of \texttt{21cmFAST} that among other useful features, enables the user to compute all of the \texttt{21cmFAST} summary statistics, while consistently accounting for an exotic dark matter model such as FDM. \texttt{21cmFAST} is a semi-numerical simulation, that enables a relatively quick generation of cosmic dawn and EoR quantities such as the baryon and dark matter density fields; the gas, spin and 21cm brightness temperature fluctuations; the neutral hydrogen fraction; and the UVLFs at various redshifts. It can account for various radiation fields, heating mechanisms and galaxy populations when computing the intergalactic medium (IGM) evolution.  
 In the case of FDM, \texttt{21cmFirstCLASS} computes the matter transfer function using \texttt{AxiCLASS}~\cite{Poulin:2018dzj, Smith:2019ihp}, and then passes it on to \texttt{21cmFAST} to generate the initial conditions (more details can be found in Ref.~\cite{Flitter:2022pzf}).
 
\subsection{UV luminosity functions}\label {section_2.1}
The UVLF describes the comoving number density of galaxies per unit absolute magnitude, $\Phi_{\rm UV} \equiv \frac{dn_{gal}}{d\Muv}$, where typically, the magnitude is measured at 1500 \r{A} in the rest frame. Following Ref.~\cite{Park:2018ljd}, the UVLF can be decomposed into three quantities: (i) $
{dn}/{d\rm M_h}$, the halo mass function (HMF); (ii) $f_{\mathrm{duty}}$, the galaxy duty cycle, which accounts for inefficient galaxy formation in small halos; and (iii) ${d\rm M_h}/{d \Muv}$, the conversion from halo mass to magnitude. These then combine to give
\begin{equation}
    \label{eq: UVLF}
    \Phi_{\rm UV} = f_{\mathrm{duty}} \, \frac{dn}{d\rm M_h} \, \frac{d\rm M_h}{d \Muv}.
\end{equation} 

The UVLF in \texttt{21cmFAST} is calculated as a product of these three quantities, but here, following the approach presented in \texttt{GALLUMI} \cite{Sabti:2021xvh, Sabti:2023xwo}, we account for the measured magnitude bin width $\Delta M_\mathrm{UV}$, and for scattering in the $M_h$ to $M_\mathrm{UV}$ relation caused by the unique formation history of each galaxy. Accounting for these effects requires integrating over the magnitude bin size, and over a range of possible halo masses for each magnitude bin:

\begin{equation}
\label{eq:integrated_LF}
\begin{split}
    \hspace{0pt} \Phi_\mathrm{UV}(M_\mathrm{UV}) = \frac{1}{\Delta M_\mathrm{UV}}\int\limits_0^\infty \mathrm{d}M_\mathrm{h} \times \\
    \hspace{0pt} \left[\frac{\mathrm{d}n_\mathrm{h}}{\mathrm{d}M_\mathrm{h}}\, f_{\mathrm{duty}} (M_\mathrm{h}) \int\limits_{M_\mathrm{UV} - \frac{\Delta M_\mathrm{UV}}{2}}^{M_\mathrm{UV}+\frac{\Delta M_\mathrm{UV}}{2}}\mathrm{d}M_\mathrm{UV}'P(M_\mathrm{UV}', M_\mathrm{h})\right] \, .
    \end{split}
\end{equation}

We now elaborate on the computation of each of the components that goes into Eq.\eqref{eq:integrated_LF}: (i) In our calculations, we use the Sheth-Tormen HMF \cite{Sheth:1999mn, Sheth:2001dp}

\vspace*{-0.1cm}
\begin{align}
    \label{eq:HMF}
    \frac{\mathrm{d}n}{\mathrm{d}M_\mathrm{h}} = \frac{\overline{\rho}_\mathrm{m}}{M_\mathrm{h}}\frac{\mathrm{d}\ln\sigma_{M_\mathrm{h}}^{-1}}{\mathrm{d}M_\mathrm{h}}f_{\rm ST}(\sigma_{M_\mathrm{h}})\ ,
\end{align}
where  $\sigma_{M_\mathrm{h}}$ is the standard deviation of the density field, smoothed over a mass scale $M_\mathrm{h}$, $\overline{\rho}_\mathrm{m}$ is the average matter energy density and the function $f_{\rm ST}(\sigma_{M_\mathrm{h}})$ is defined as 

 \begin{align}
    \label{eq:ST_HMF}
    f_\mathrm{ST}(\sigma_{M_\mathrm{h}}) = & \rm A\sqrt{\frac{2a}{\pi}}\left[1+\left(\frac{\sigma_{M_\mathrm{h}}^2}{\delta_c^2}\right)^{p}\right]\frac{\delta_c}{\sigma_{M_\mathrm{h}}}\exp\left(-\frac{a\delta_c^2}{2\sigma_{M_\mathrm{h}}^2}\right),
\end{align}
%\enlargethispage{0.2cm}
where the values of the free parameters $A=0.3222$, $a= 0.707$ and $p=0.3$ were fit using numerical simulations, and $\delta_c = 1.686$ is the linear density field critical collapse threshold.

The effect of FDM on our model of the UVLFs enters through the computation of   
$\sigma_{M_\mathrm{h}}$, which can obtained by

\begin{align}
    \label{eq:sigmasq_M}
    \sigma^2_{M_\mathrm{h}}(z) &= \int\frac{\mathrm{d}^3k}{(2\pi)^3}W_{M_\mathrm{h}}^2(k)T^2_\zeta(k,z)P_\zeta(k)\ ,
\end{align}
where $W_{M_\mathrm{h}}$ is the Fourier transform of the spherical top hat filter, $T_\zeta(k,z)$ in the matter transfer function, obtained using \texttt{AxiCLASS} as we mentioned earlier, and $P_\zeta(k)$ is the primordial power spectrum. Although Refs.~\cite{Benson:2012su,Kulkarni:2020pnb} suggest that a top-hat filter may not be appropriate for a matter power spectrum with a cutoff as it may lead to an overestimation of the number of small halos, in our analysis we mostly focus on small FDM fractions ($f_\mathrm{FDM}\lesssim10\%$, see Fig.~\ref{fig:all_constraints}), and thus the HMF in our simulations behaves more similarly as in $\Lambda$CDM. In any case, this is a conservative choice and using a different k-filter may lead to slightly stronger constraints.
The matter transfer function depends on the values of the cosmological parameters, $H_0$, $\Omega_m$, $\Omega_b$ and the primordial power spectrum depends on $A_s$ - the primordial fluctuations amplitude, and $n_s$ - the primordial power index.  
We set all the parameters mentioned here, except $\Omega_b$, as free parameters in our model, letting them vary within the $2\sigma$ bounds achieved by the Planck collaboration \cite{Planck:2018vyg}. $\Omega_b$ is left out at this point since reducing the dimension of the parameter space improves the emulator accuracy (see section \ref{section_3}), and $\Omega_b$ has the weakest imprint on the observables from all the cosmological parameters.

(ii) We model $f_{\mathrm{duty}}$ as an exponential cutoff at a turnover mass $\rm M_{\rm turn}$ below which galaxy formation is suppressed by feedback mechanisms, or inefficient gas accretion \cite{Shapiro:1993hn, Hui:1997dp, Barkana:2000fd, Springel:2002ux, Okamoto:2008sn, 
Mesinger:2008ze,
Sobacchi:2014rua, 
Sobacchi:2015gpa}, 
\begin{equation}\label{f_duty}
    f_{\mathrm{duty}}(M_{\rm h} ) = {\rm exp}(-\frac{M_{\rm turn}}{M_{\rm h}} ).
\end{equation}

(iii) Finally, the probability density of the UV magnitude, $P(M_\mathrm{UV},M_\mathrm{h})$, is modeled as a Gaussian distribution with width $\sigma_\mathrm{UV}$ and mean $\overline{M_\mathrm{UV}}$. 
In order to evaluate  $\overline{M_\mathrm{UV}}(M_\mathrm{h})$, we first note that the AB magnitude is related to the UV luminosity  through \cite{Oke:1983nt},
\begin{align}
    \label{eq:LUV}
    \log_{10}\left(\frac{L_\mathrm{UV}}{\mathrm{erg \, s^{-1}}}\right) = 0.4\, (51.63 - M_\mathrm{UV})\ .
\end{align}
The UV luminosity in turn is assumed to be proportional to the star formation rate (SFR) via the conversion factor $\kappa_\mathrm{UV} = 1.15\times 10^{-28}\, M_\odot\mathrm{\,s\, erg}^{-1}\mathrm{yr}^{-1}$
\cite{Madau:1997pg, Kennicutt:1998zb}
such that 

\begin{align}
    \label{eq:Mstardot_LUV}
    \dot{M}_* = \kappa_\mathrm{UV}L_\mathrm{UV}\ .
\end{align}
The average SFR is modeled as the total stellar mass, $M_\ast(M_\mathrm{h})$, divided by a characteristic time scale,

\begin{equation}\label{eq: SFR}
\dot{M_{\ast}}(M_{\rm h},z) =  \frac{M_\ast(M_\mathrm{h})}{t_\ast H(z)^{-1}},
\end{equation}

where $H(z)$ is the Hubble parameter, and the star formation timescale $t_{\ast}$ is a model parameter that can take a value between 0 and 1. Following former works we set \cite{HERA:2021noe} $t_{\ast}$ =0.5.

For the total stellar mass, rather than using the default  prescription in \texttt{21cmFAST}, which is not designed to be computed at the bright end of the observed luminosity function (the stellar to halo mass ratio can reach unity for massive halos that populate the bright end), we modify the \texttt{21cmFAST} code and implement instead  the prescription introduced in \texttt{GALLUMI}~\cite{Sabti:2021xvh, Sabti:2023xwo}, 
\begin{align}
    \label{eq:f_star}
    M_\ast = \dfrac{\epsilon_*}{\left(\dfrac{M_\mathrm{h}}{M_p}\right)^{\alpha_*}+\left(\dfrac{M_\mathrm{h}}{M_p}\right)^{\beta_*}} M_\mathrm{h}\ ,
\end{align}
where $\alpha_* > 0$, $\beta_* < 0$, and $\epsilon_\ast$, $M_c$ are redshift dependent amplitude and pivot mass respectively, with redshift pivoting at $z=6$, 
$ \log_{10}\epsilon_*(z) = \epsilon_*^\mathrm{i} + \epsilon_*^\mathrm{s}\times \log_{10}\left(\frac{1+z}{1+6}\right)$, $
        \log_{10}\frac{M_p(z)}{M_\odot} =  M_p^\mathrm{i} + M_p^\mathrm{s}\times \log_{10}\left(\frac{1+z}{1+6}\right)$.
The major difference between the \texttt{21cmFAST} and \texttt{GALLUMI} parameterizations, is that \texttt{GALLUMI}'s involves a double power law with different slopes for the bright and faint ends of the luminosity function. On the other hand the \texttt{21cmFAST} model has a single power law and is thus restricted to a single slope for all masses. We will use our modified \texttt{21cmFAST} prescription for all the calculations involving the SFR.

\begin{figure*}[ht!]
	\centering
	\includegraphics[width = \columnwidth]{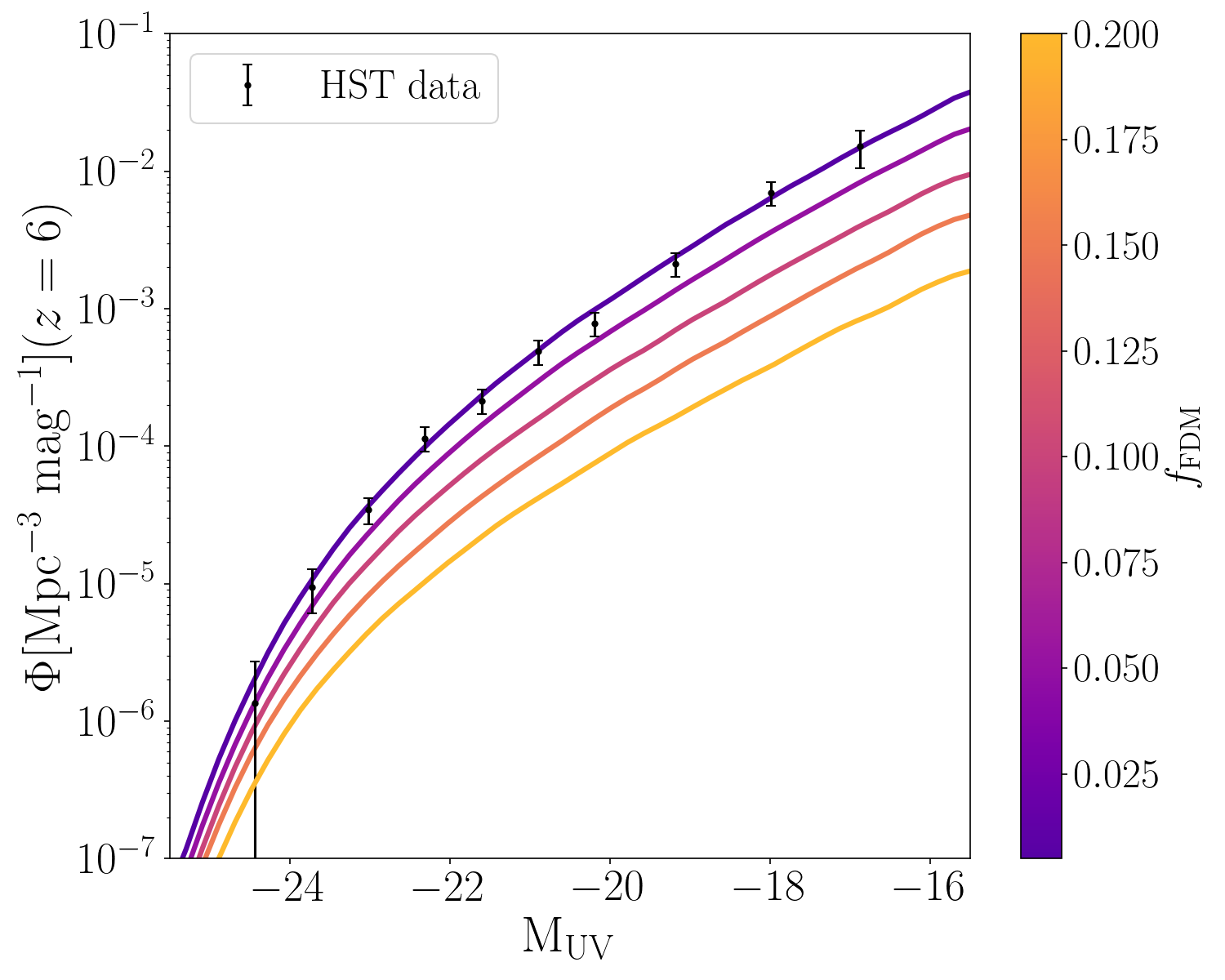}
	\includegraphics[width = \columnwidth]{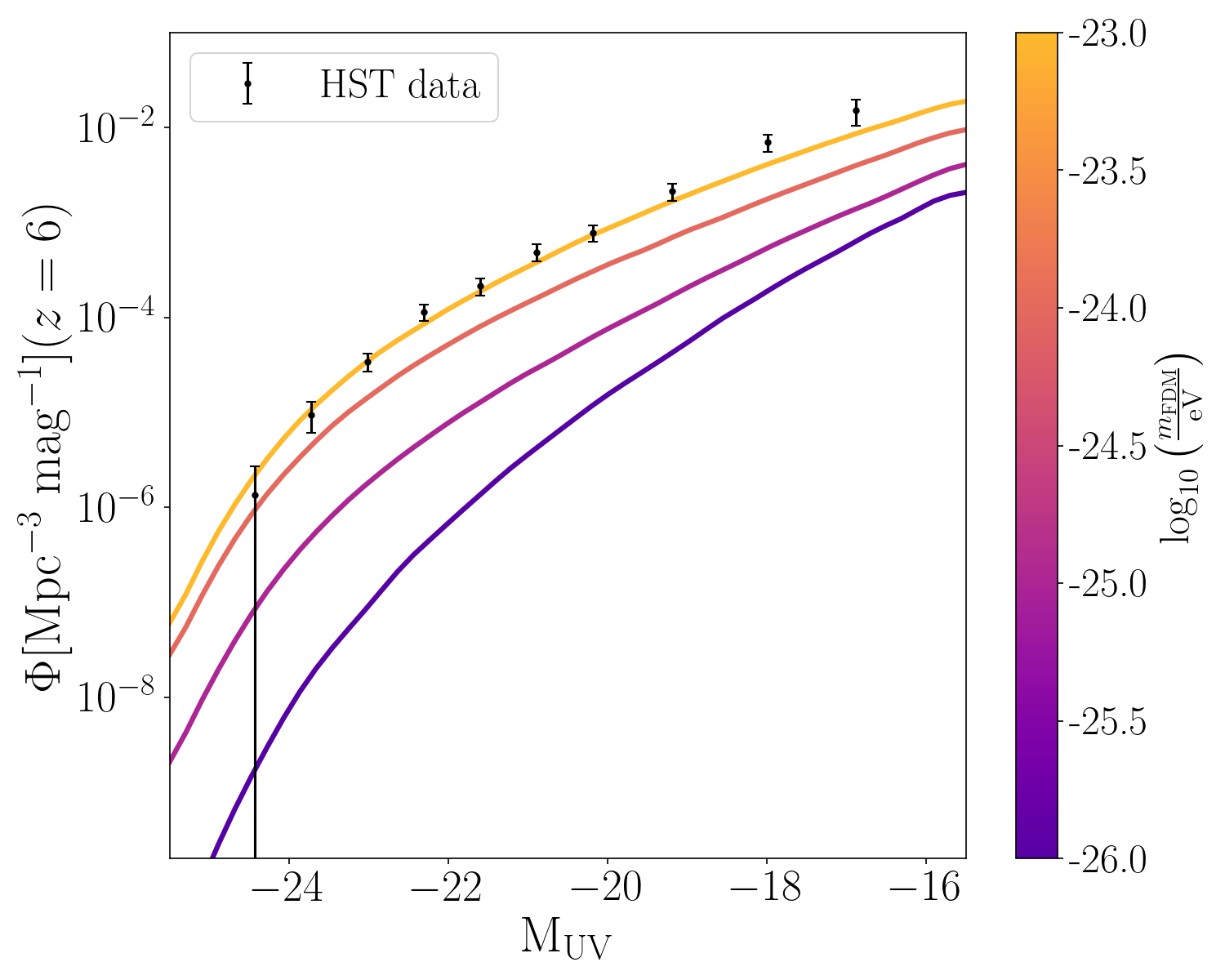}
	\caption{The HST UVLFs at redshift $z=6$. {\it Left:} Here we set $m_{\rm FDM} = 10^{-24} {\rm eV}$ and vary $f_{\rm FDM}$, taking the median values for the UVLF model parameters from the MCMC posterior distribution shown further below. Increasing the FDM fraction decreases the UVLF amplitude, with a stronger effect on the faint end of the UVLF. {\it Right:} Here we set $f_{\rm FDM} = 0.1$ and vary $m_{\rm FDM}$. Decreasing the FDM mass decreases the UVLF amplitude, with a stronger effect on the bright end of the UVLF.
 }
\label{fig:LF_z=6_M=-23}
\end{figure*}

The effect of different values in the FDM parameter space on the UVLFs is demonstrated in Fig.~\ref{fig:LF_z=6_M=-23}, alongside the HST measurements. It is clear from this figure that the small scale suppression of FDM has a major effect on the UVLFs, and has the possibility to push it beyond the observed limits. The left panel indicates that at a fixed FDM mass, increasing the FDM fraction has a stronger effect on lower luminosities (higher magnitude). The right panel shows that decreasing the FDM mass for fixed  FDM fraction constant decreases the galaxy number density with stronger impact at high luminosity.

Before we proceed to execute the parameter inference, the UVLF HST measurements need to be modified by taking into account a few physical effects that can impact the relation between the physical UVLF and the measured one. All of the data corrections listed here are implemented in the same manner as in \texttt{GALLUMI}~\cite{Sabti:2021xvh}:

(i) Cosmic variance - as the sky patches covered by HST at high redshifts are relatively small \cite{Trapp:2020cde},
the fluctuations in the matter density field can lead to a bias in the measured LFs. To account for this, we impose a conservative minimal error of 20\% on all the data points.

(ii) Dust attenuation - the UV flux of a galaxy can be absorbed by dust inhabiting the interstellar medium (ISM), while the outgoing emission from the dust is in the infrared. This effect can bias downwards the measured luminosity. Since dust accumulates mostly in massive galaxies, the impact is more dominant on the bright end of the LFs, at low redshifts. The dust extinction can be modeled using the IRX-$\beta$ relation~\cite{Meurer:1999jj}; our implementation follows Ref.~\cite{Sabti:2021xvh} (see there for further details).

(iii) Alcock-Paczynski effect - the definition of the luminosity function is the number of galaxies in a magnitude bin, divided by the survey volume, where some cosmological model and parameters are assumed when calculating the volume.
This implies that when one varies the cosmological parameters, or differs from $\Lambda{\rm CDM}$ Cosmology, the data points and their errors should be reevaluated~\cite{Alcock:1979mp}. In our case, FDM does not affect the cosmological volume, but the variation of the cosmological parameters does affect the inference.

\subsection{Neutral  fraction and optical depth}\label{section:2_2}

The average Hydrogen neutral fraction, $\Bar{x}_{\rm HI}$, at redshifts $5-35$, is one of the \texttt{21cmFAST} summary statistics, and as such, it can also be obtained using \texttt{21cmFirstCLASS} (at even higher redshifts). The simulated value of $\Bar{x}_{\rm HI}$ is influenced by the SFR parameters, defined in Eq.~\ref{eq:f_star}, but also by the galaxies UV escape fraction, which is parameterized similarly to the original \texttt{21cmFAST} stellar to halo mass ratio, as a power law

\begin{equation}\label{eq:F_ESC}
f_{\rm esc}(M_{\rm h}) = f_{{\rm esc, 10}}\left( \frac{M_{\rm h}}{10^{10}{\rm M}_{\odot}}\right)^{\alpha_{\rm esc}},
\end{equation}
where the normalization factor $f_{{\rm esc, 10}}$ and the power law index $\alpha_{\rm esc}$ are free parameters.

Following Ref.~\cite{Shmueli:2023box}, we calculate the optical depth to reionization analytically, using the output from \texttt{21cmFirstCLASS}: $x_{\rm HI}$ mentioned above, and  the baryon  density field $\delta_b$. It is important to note that we do not use the simulation-averaged quantities, but the value at each simulated cell. Here, we skip some parts of the derivation and present the final form for the optical depth (for more details the reader is referred to, e.g., Refs.~\cite{Shmueli:2023box, Liu:2015txa}):

\begin{eqnarray}
\label{eq:tau}
\tau = \frac{3 H_0 \Omega_b \sigma_Tc}{8 \pi G m_p} \left[ 1 + \frac{Y_p^\textrm{BBN}}{4}\left( \frac{m_\textrm{He}}{m_\textrm{H}} - 1\right)\right]^{-1} \nonumber \\ 
\times \int_0^{z_\textrm{CMB}} \frac{dz (1+z)^2}{\sqrt{\Omega_\Lambda + \Omega_m (1+z)^3}}  \overline{x_\textrm{HII} (1+\delta_b)}
\end{eqnarray}
where $\sigma_T$ is the Thompson cross section, $m_i$ is the i'th species mass, $Y_p^\textrm{BBN} = n_\textrm{He} / n_b $ is the Helium fraction, and $x_{\rm HII} = 1 - x_{\rm HI}$ is the ionized fraction. Since \texttt{21cmFAST} simulates the evolution of the IGM down to redshift $z=5$, the integral in Eq.~\ref{eq:tau} cannot be evaluated directly for $z<5$. In order to calculate it, we assume that at redshift $z=5$ the universe in already completely ionized, which implies that $\overline{x_{\rm HII}(1+\delta_b)} \approx 1$. This is a reasonable assumption, since the neutral fraction is already constrained to be less than $0.11$ at $z=5.9$ \cite{McGreer:2014qwa}, however, in rare scenarios, this calculation is somewhat inaccurate. Such a scenario can be caused for significant fraction of FDM with a very small mass, but this will not bias our inference, since the optical depth will already be too small to fit the Planck~\cite{Planck:2018vyg} CMB bounds on $
\tau$.

\begin{figure*} [t!]
	\centering
	\includegraphics[width = \columnwidth]{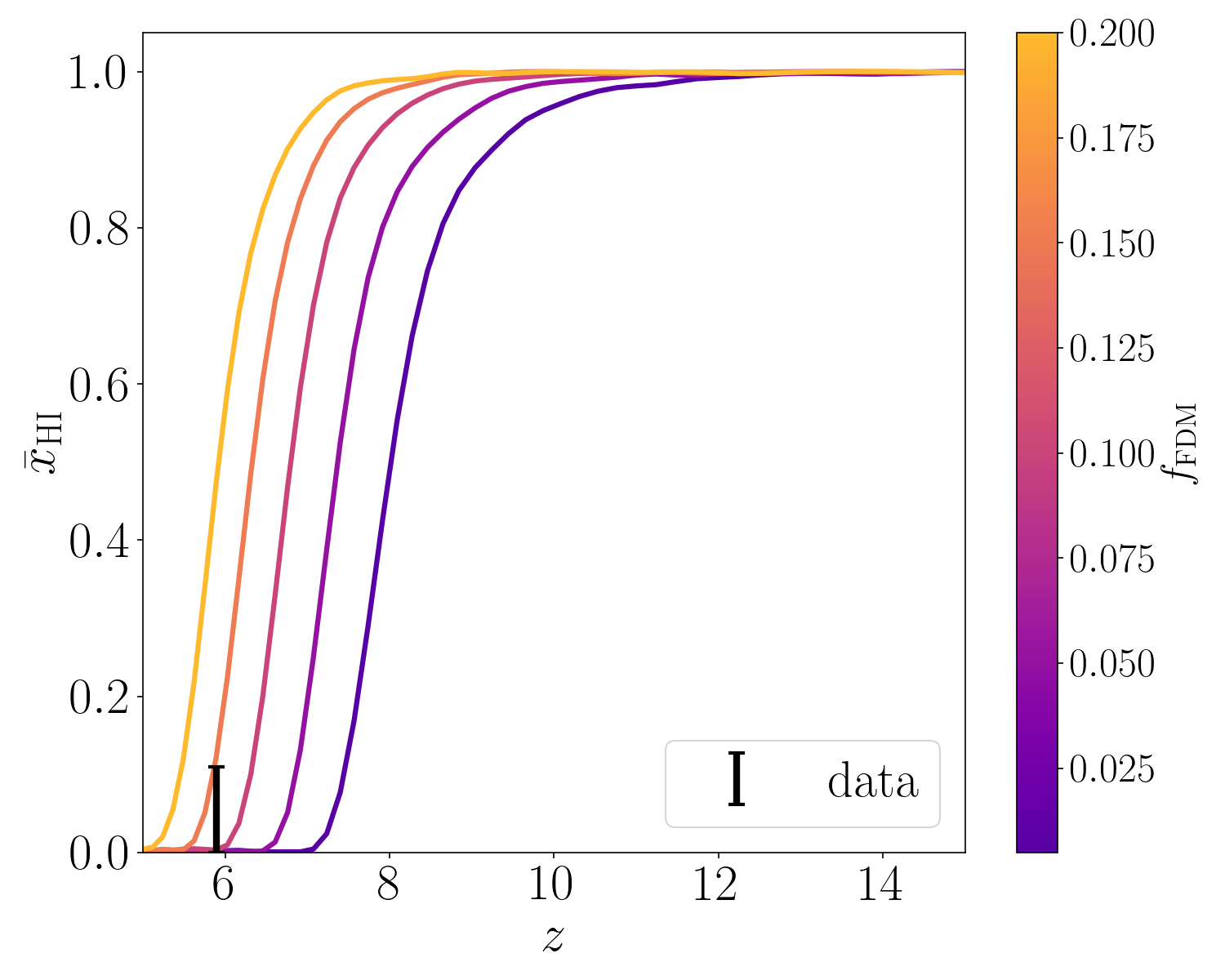}
	\includegraphics[width = \columnwidth]{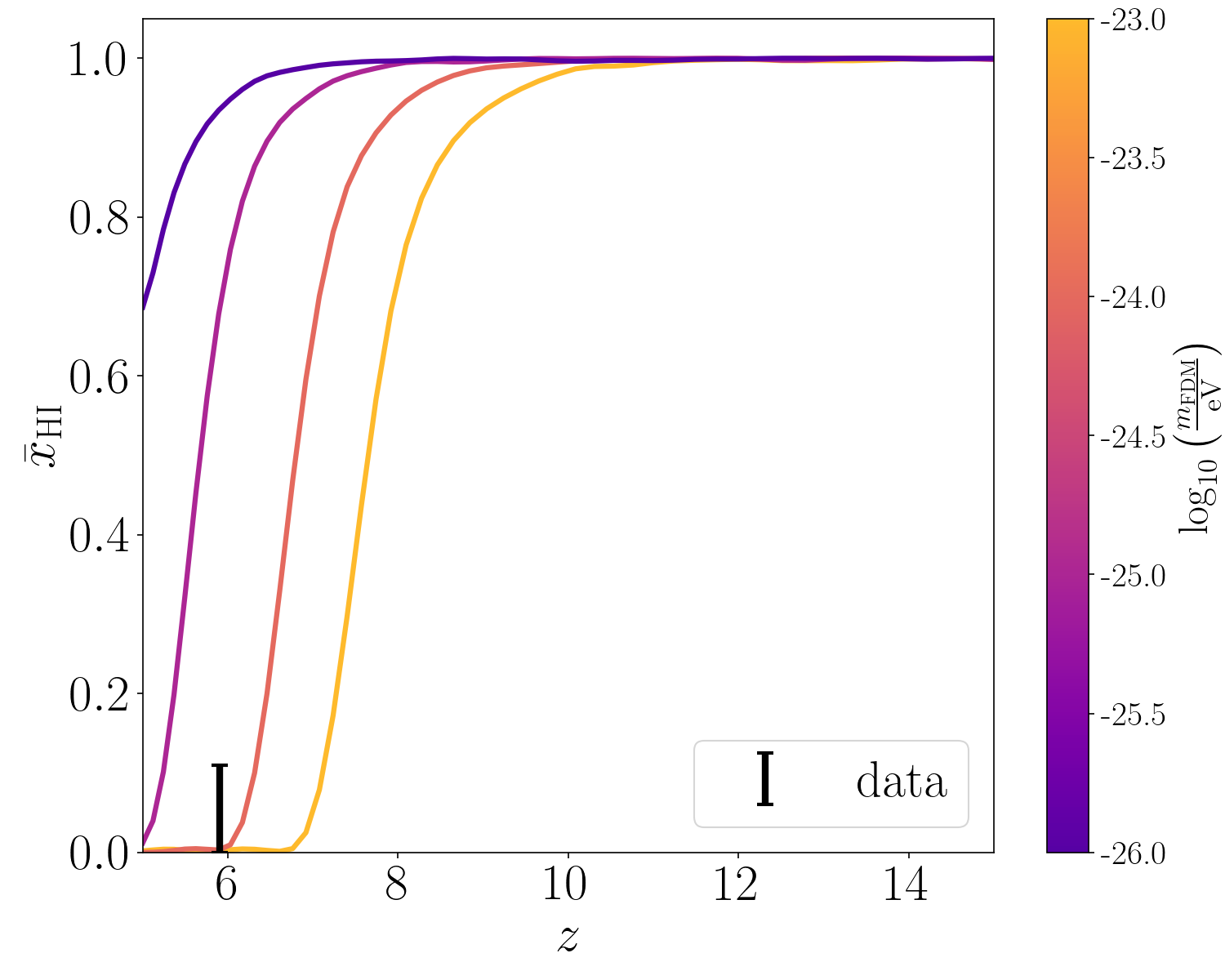}
	\caption{The neutral hydrogen fraction redshift evolution when varying the FDM fraction or mass, calculated using our emulator. The datapoint is $x_{\rm HI}<0.11$ at $z=5.9$ \cite{McGreer:2014qwa}. {\it Left:} Here we take $m_{\rm FDM} = 10^{-24} \, {\rm eV}$ and set the model parameters to their median values from the MCMC posterior distribution shown below. The effect of  FDM is to delay structure formation which in turn translates into a delayed reionization. Large fractions delay reionization beyond the current limits. {\it Right:} Varying the mass, while fixing $f_{\rm FDM} = 0.1$.  We see that small masses delay reionization beyond  known limits, even for small FDM  fractions.}
\label{fig:x_HI_varing_f_FDM}
\end{figure*}

\begin{figure} [h]
	\centering
	\includegraphics[width = \columnwidth]{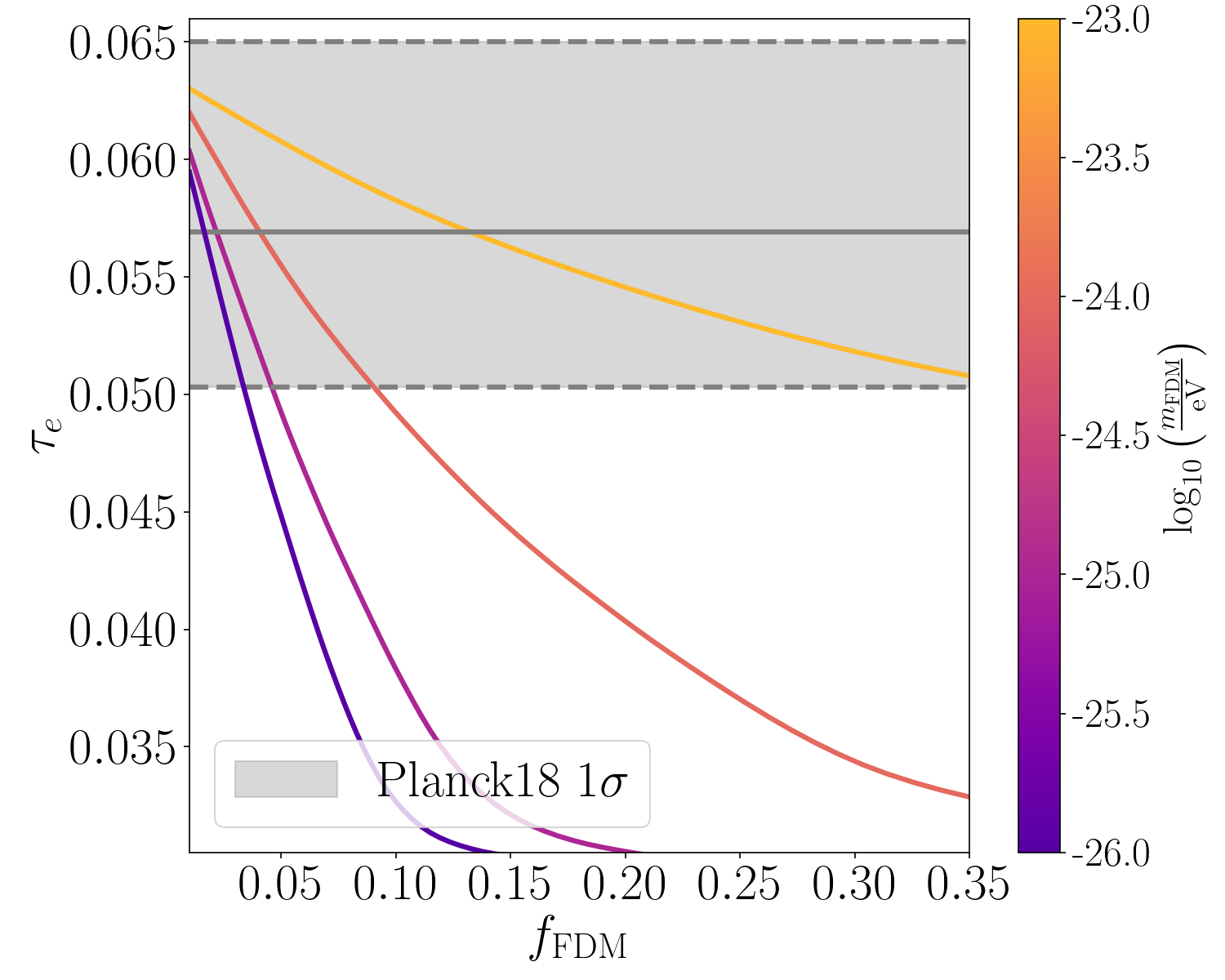}
	\caption{The Thomson scattering optical depth to reionization when varying the FDM fraction, achieved using an emulator. The figure was obtained median parameters from the MCMC posterior distribution shown below. Here the effect of FDM is delaying structure formation which in turn translates into a delayed reionization, and a larger neutral hydrogen fraction at low redshifts. Eq.~\ref{eq:tau} implies that $\tau_e \propto 1-x_{\rm HI}$, which means that a larger a neutral fraction at low redshifts, will result in a decreasing optical depth.  }
\label{fig:tau_varing_f_FDM}
\end{figure}

In Figs.~\ref{fig:x_HI_varing_f_FDM},\ref{fig:tau_varing_f_FDM} we demonstrate the implications of FDM on $\tau_e$ and $x_{\rm HI}$. In a similar manner to the effects on the UVLFs, we see here that FDM has noticeable imprints on these observables, and can be constrained using them.
The source of these imprints is the small scale suppression that FDM generates at early times. This suppression delays the formation of heavy DM halos, which in turn delays the formation of galaxies. The UV flux from those galaxies is responsible for the reionization of the IGM, which means that this delay can be measured using the neutral hydrogen fraction~\cite{McGreer:2014qwa,Spina:2024uyc}, and with the optical depth to reionization.

\subsection{The 21cm power spectrum} 

Emission and absorption of the 21cm line in the IGM are characterized by the spin temperature, $T_s$, which models the population ratio between the two hyperfine states.
It is usually measured as the differential brightness temperature, $\delta T_{21}$, 
with respect to the brightness temperature of the low-frequency radio background, $T_{\rm{rad}}$.
In the standard  scenario, $T_{\rm{rad}}$ is taken to be the CMB temperature $T_{\rm{cmb}}=2.7254 \times (1+z)$ K.
The differential brightness temperature, $\delta T_{21}$, is given by \cite{Bharadwaj:2004nr, Furlanetto:2009iv, Madau:2015cga} 
\begin{equation}\label{eq:21cm_signal}
    \begin{aligned}
        \delta T_{21}(\nu) &= \frac{T_{\mathrm{s}}-T_{\rm rad}}{1+z}\left(1-e^{-\tau_{\nu_{0}}}\right) \, ,
    \end{aligned}
\end{equation}
where $\left(1-e^{-\tau_{\nu_{0}}}\right)$ takes into account the effect of propagation 
through a medium, and $\tau_{\nu_{0}}$ is the optical depth. Most of the astrophysical and cosmological 
information is captured in $T_{\mathrm{s}}$, whereas the state of the medium 
is encoded in the optical depth $\tau_{\nu_{0}}$. 
Measurements of the 21cm signal include both the sky direction averaged signal $\langle \delta T_{21}(z) \rangle$, known as the global signal, and its fluctuations that can be interpreted in terms of  
statistics like the power spectrum, 
bispectrum, etc. Here, we work with the power spectrum defined as 
\begin{equation}
\label{eq:power_spectra}
    \langle \tilde{\delta T}_{21}({\bf k}_1) \tilde{\delta T}_{21}({\bf k}_2) \rangle = (2 \pi)^3 \delta^D({\bf k}_1 - {\bf k}_2) P_{21}({\bf k}_1)\,,
\end{equation}
where $\langle...\rangle$ denotes an ensemble average, $\tilde{\delta T}_{21}({\bf k})$ is the  Fourier transform of $ \delta T_{21}(\bf x)$ and $\delta^D$ is the Dirac delta function. Specifically, the quantity of interest for us is $\Delta_{21}^2({\bf k}) \equiv k^3 P_{21}({\bf k})/(2 \pi^2)$, which is the output of HERA \cite{HERA:2022wmy} measurements. Needless to say, realizations of the 21cm global signal and spatial fluctuations can also be generated using \texttt{21cmFirstCLASS}, where the SFR and escape fraction parameters affect the output signal.

\begin{figure*} [t!]
	\centering
	\includegraphics[width = \columnwidth]{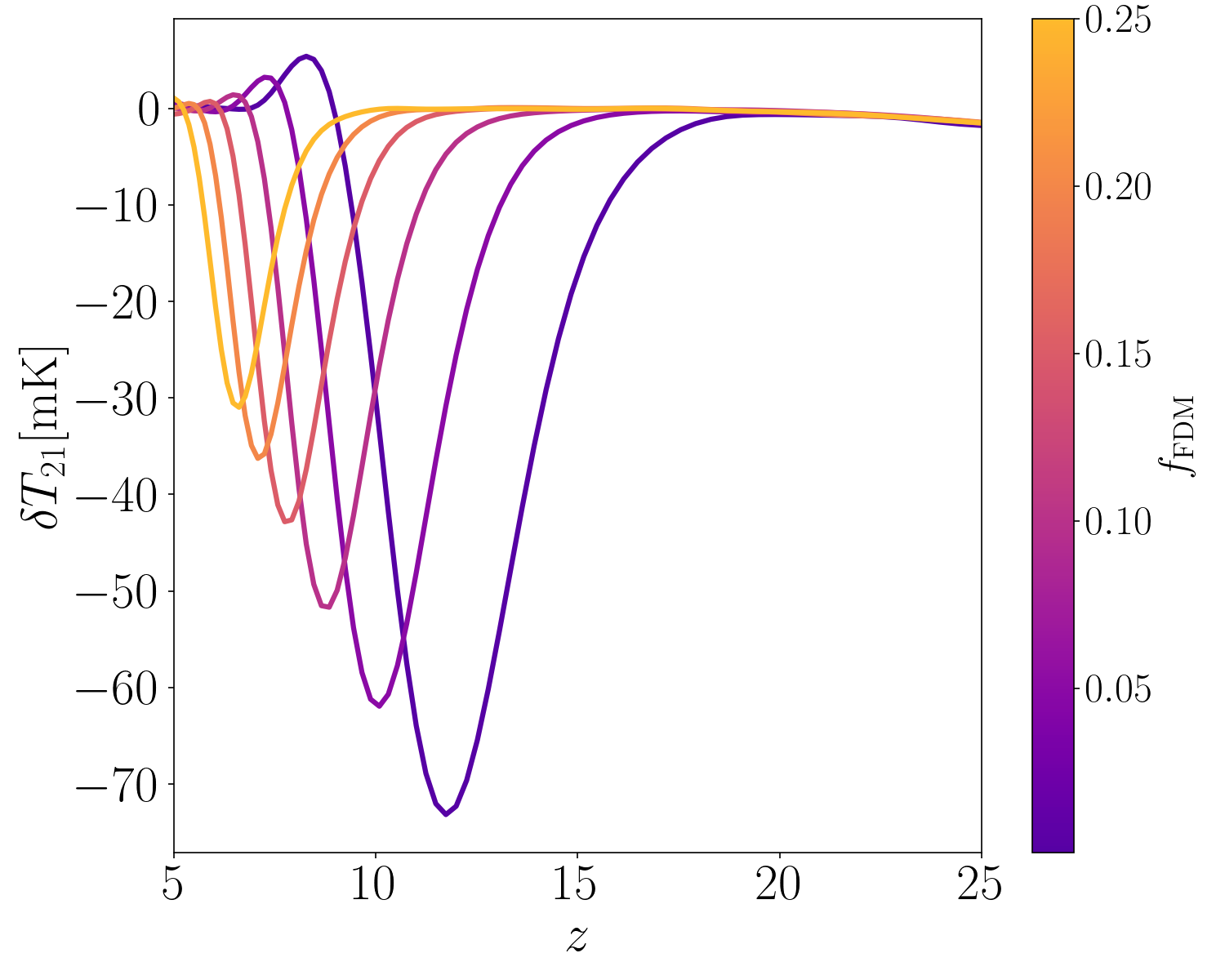}
	\includegraphics[width = \columnwidth]{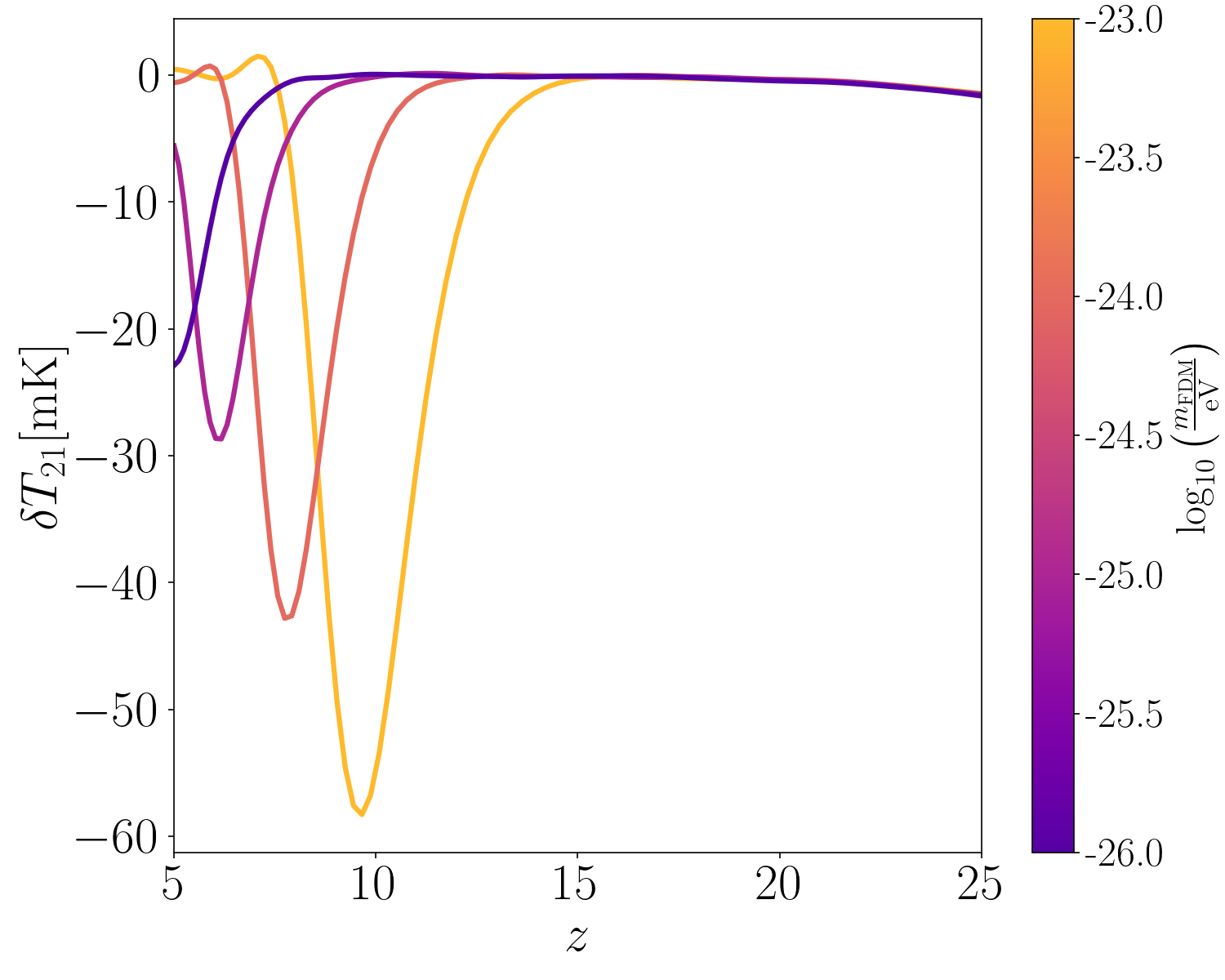}
	\caption{The 21cm global signal redshift evolution when varying the FDM fraction ({\it Left}) or FDM mass ({\it Right}), calculated using our emulator. The delay of structure formation delays all the phases of the 21cm signal, from cosmic dawn to reionization. {\it Left:} Here we take $m_{\rm FDM} = 10^{-24} \, {\rm eV}$ and set the global signal parameters to their median values from the MCMC posterior distribution shown below. {\it Right:} Here we fix $f_{\rm FDM} = 0.15$.}
\label{fig:T21_varing_f_FDM}
\end{figure*}

As demonstrated in Ref.~\cite{Flitter:2023mjj}, the delay in structure formation that FDM incurs, will result in a delay in the 21cm signal. This delay is clearly demonstrated by the global signal as shown in Fig.~\ref{fig:T21_varing_f_FDM}. Since the global signal experiments \cite{Singh:2021mxo, Monsalve:2017mli, Monsalve:2018fno, Monsalve:2019baw} are not yet sensitive enough to detect this effect, we can try to use the 21cm power spectrum measurement by HERA \cite{HERA:2022wmy}. The results of this experiment are currently treated as upper bounds on the power spectra, since the systematic noise is not yet modeled reliably \cite{HERA:2021bsv, HERA:2021noe}. The likelihood for the HERA upper bounds can be found in Refs \cite{HERA:2021bsv, Lazare:2023jkg}.
In practice, it turns out that the current upper bounds do not have any constraining power over the model parameters we use in this work, since all of them are already bound by the other observables. In section \ref{section_4} this will be demonstrated for a specific
FDM mass. Future power spectrum detection by HERA should improve the constraints achieved in this work. In Section~\ref{section 4.2} this will be demonstrated using a mock observation of the power spectra, assuming a future HERA noise specification.

\section{Artificial Neural Network (ANN)}\label{section_3}

Since the semi-numerical computation of $\tau_e$, $x_{\mathrm{HI}}$ and $\Delta_{21}^2 (k,z)$ requires running a complete \texttt{21cmFirstCLASS} simulation, the computational time for each set of parameters, can take up to O(1 hour) even on a high performance computing environment (HPC). The traditional  MCMC pipeline requires generating a large number of simulations in a sequential manner in order to construct the final posterior. This can result in a few weeks for each inference, and can be prohibitive, and not scalable. In order to address this problem, we turn to use the developing field of artificial neural networks (ANNs) to construct an EoR summaries emulator which we describe below. Our work here is based on our previous work presented in Ref.~\cite{Lazare:2023jkg}, and on the 21cm emulator \texttt{21cmEMU}~\cite{Breitman:2023pcj}.

\subsection{Building the dataset}
In this section, we discuss the method of composing a dataset which will then be used to train and test our emulator.
Training a reliable emulator in a high dimensional parameter space, can require a very large number of simulations. For that reason we set some of the model parameters to a constant value, determined by a prior MCMC run with UVLFs only, and without FDM. Such MCMC does not require an emulator since the realizations can be produced rapidly. The  parameters we set to constants are: $\epsilon_\ast^s=0.82$, $M_p^s = 2.37$ and $\sigma_\mathrm{UV}=0.45$. We also set $\Omega_b$ to the best-fit value from Planck~\cite{Planck:2018vyg}.

The free parameters and their ranges are summarized in Table~\ref{tab:prior_table}. Next, we sample the parameter space using a Latin hypercube (LH) sampler which aims to produce uniform sampling when all points are marginalized onto any one dimension \cite{McKay1979}. In this work, we are training our emulator, without varying the axion mass. This means that we  have to train a separate emulator on a separate data set, for each axion mass. We examine 7 different values for the axion mass varying from $10^{-23} \,{\rm eV}$ to $10^{-26} \, {\rm eV}$, and for each of them we generate $\sim\!10000$ different parameter sets.
Then, for each of the parameter sets obtained, we compute the luminosity functions, the neutral hydrogen fraction, the optical depth to reionization, and the 21cm global signal and power spectrum using \texttt{21cmFirstCLASS}, and save them together. The UV luminosity function is then interpolated over 50 magnitude bins, ranging from -25.5 to -15.5 magnitudes. Finally, the samples are split into a training set (10\%), validation set (10\%) and testing set (80\%). Before the
training process begins, the parameter sets are normalized to lie in the range [-1, 1], 
and we use the $\log_{10}$ of the corresponding UVLFs and 21cm power spectra to train the ANN, as former publications had suggested that this accelerates the learning process and increases accuracy.

\begin{figure*}
    \centering
\includegraphics[width = 0.8\textwidth]{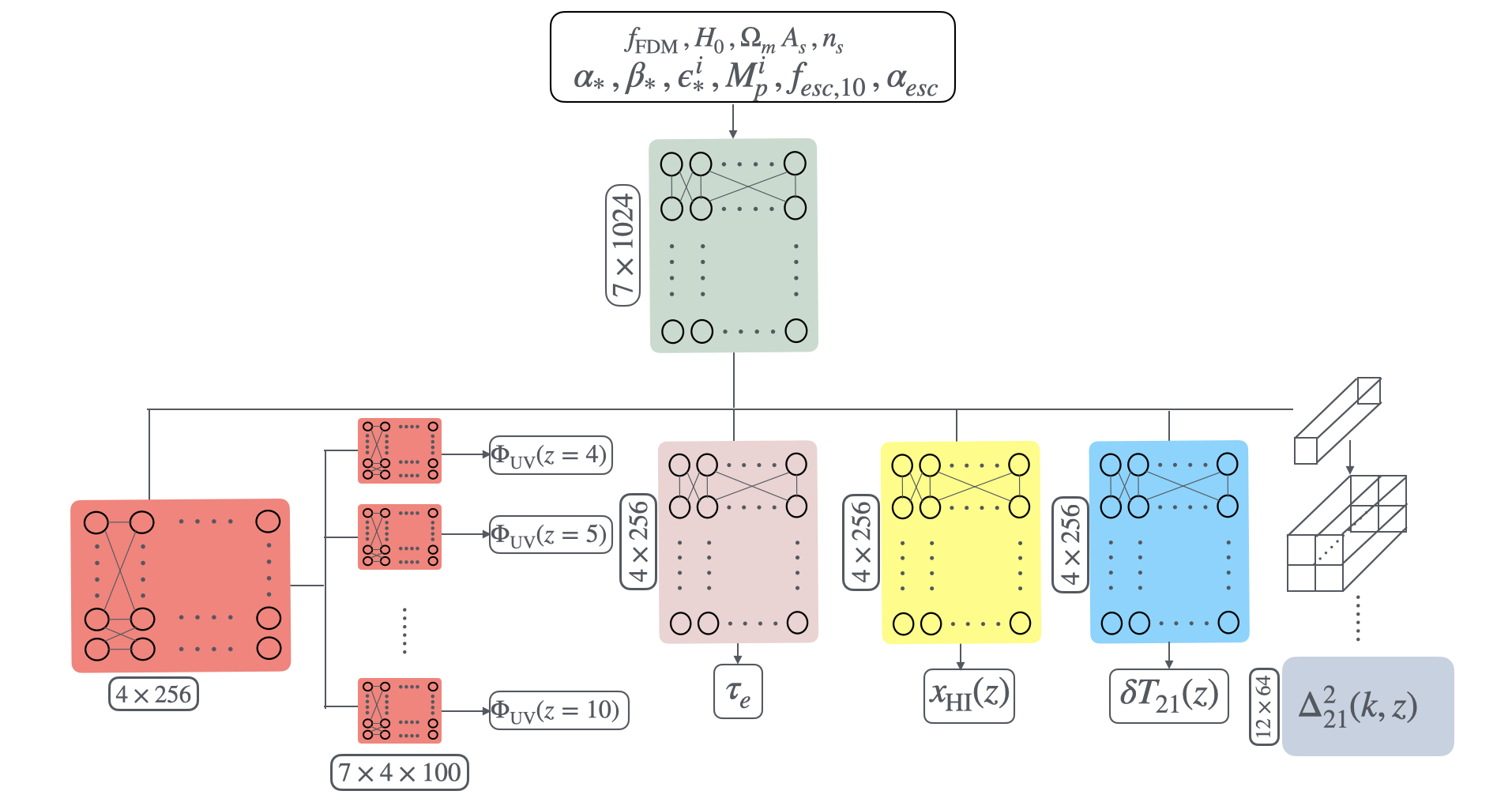}\\
\vspace{-0.2in}
\caption{A schematic diagram of the emulator architecture. The  astrophysical and cosmological parameters are inserted to a large block of fully-connected layers. The output from this shared block is then passed on to four fully connected blocks, and one convolutional block for the 21cm power spectrum. 
The pink block predicts the Thomson scattering optical depth, the yellow block is responsible for the neutral fraction, the blue block is for the 21cm global signal, and the red block is a shared block for the luminosity functions. The output of the shared UVLF block is then passed on to seven smaller fully connected blocks, each predicts the luminosity function at a different redshift.
}\label{fig:schematic_NN}
\end{figure*}

\subsection{ANN architecture}

A schematic diagram of the emulator in shown in Fig.~\ref{fig:schematic_NN}. The emulator is implemented using the \texttt{TensorFlow}~\cite{tensorflow2015-whitepaper} and \texttt{Keras}~\cite{chollet2015keras} libraries defined in \textsc{Python3}. For all of the emulator outputs except the 21cm power spectra, we use fully connected (Dense) layers, where the number of layers and the number neurons in each layer are specified in Fig.~\ref{fig:schematic_NN}. For the power spectrum, we exploit the power of convolutional neural networks (CNNs)~\cite{ciresan2011flexibles, Szegedy:2014nrf}, and use convolution layers, together with transpose convolution layers, and up-sampling layers, that reshape the output until the required form is achieved. 
For each of the predicted observables, we use a mean squared error (MSE) loss function, to model the difference between the predictions for the Ground Truth (GT) data, and an exponential linear unit (ELU) \cite{Clevert:2015qwe} activation function, in order to insert non-linearity into the network. We have examined different conventional choices for the activation, such as ReLU~\cite{4082265}, LeakyReLU, tanh and more, and concluded that ELU delivers the most accurate results.
The training process is performed using mini batches of size 256 samples, so that the trainable parameters are updated using the gradients of the loss function, after forward propagating each batch. The training process was done using the Adam \cite{Kingma:2014vow} optimizer, where the initial learning rate is 0.001, and is reduced by  a factor of 2 when the prediction on the validation set does not improve over more than 5 epochs. The training phase ends when there is no improvement over more than 15 epochs, and the best weights are stored at that moment.

\subsection{ANN performance}
The emulator performance is tested on a 1000 sample test set, which was generated together with the training and validation samples. The metric we choose to examine is the fractional error (FE) defined for each emulator prediction as 
\begin{equation}\label{FE}
    {\rm FE}(y_{\rm pred}) = \frac{1}{N_{\rm bins}}\sum_{n=1}^{N_{\rm bins}}\frac{|y_{\rm true}(n) - y_{pred}(n)|}{y_{\rm true}(n)} \times 100 ,
\end{equation}
 where $N_{\rm bins}$ is the number of bins (redshift for $x_{\rm HI}$ magnitude for $\Phi_{\rm UV}$ etc.) at which we measure the observable.
 It is important to note that the evaluation of the FE metric for $x_{\rm HI}$ and $\Delta_{21}^2(k,z)$ is not possible naively, since their value can go down to zero for some sets of parameters.
For that reason, we force a lower bound on the GT values of the neutral fraction, $x_{\rm HI, min} = 10^{-3}$, and the 21cm power spectra, $\log_{10} {\Delta^2_{21, min}} = 10^{-1} \, \mathrm{mk^2}$.

 \begin{figure}[h!]
	\centering
	\includegraphics[width = \columnwidth]{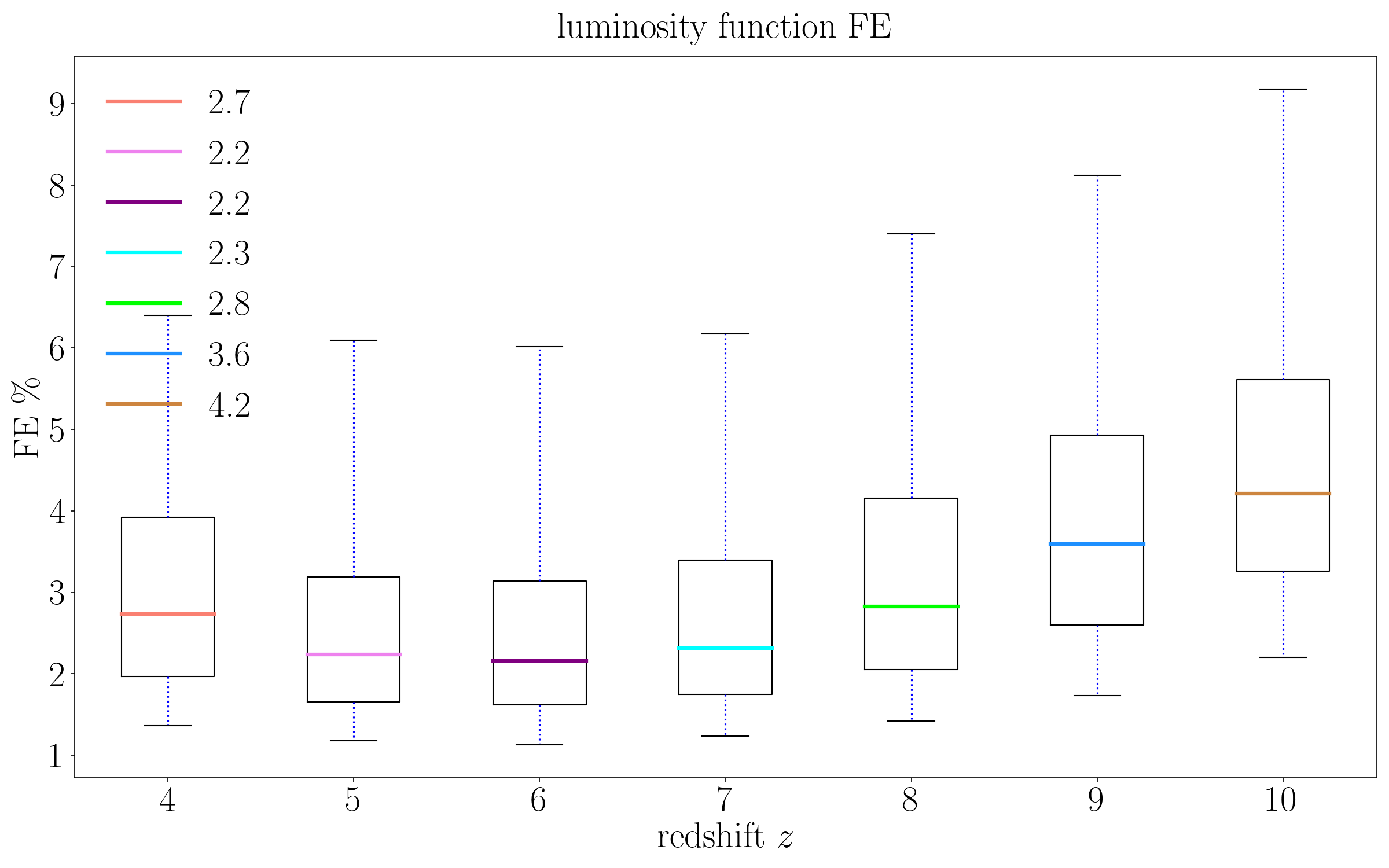}
	\caption{UVLF accuracy statistics of the testing set, for $m_{\rm FDM} = 10^{-24
} \, {\rm eV}$. The error is averaged over the magnitude bins. The boxes represents all the samples that fall between the 25th and the 75th  FE percentile, and the line in the middle of it is the median. The UVLFs errors statistics of the other FDM masses examined here, distributes similarly.}
\label{fig:UVLF_testing}
\end{figure}

\begin{figure}[h!]
	\centering
\includegraphics[width = \columnwidth]{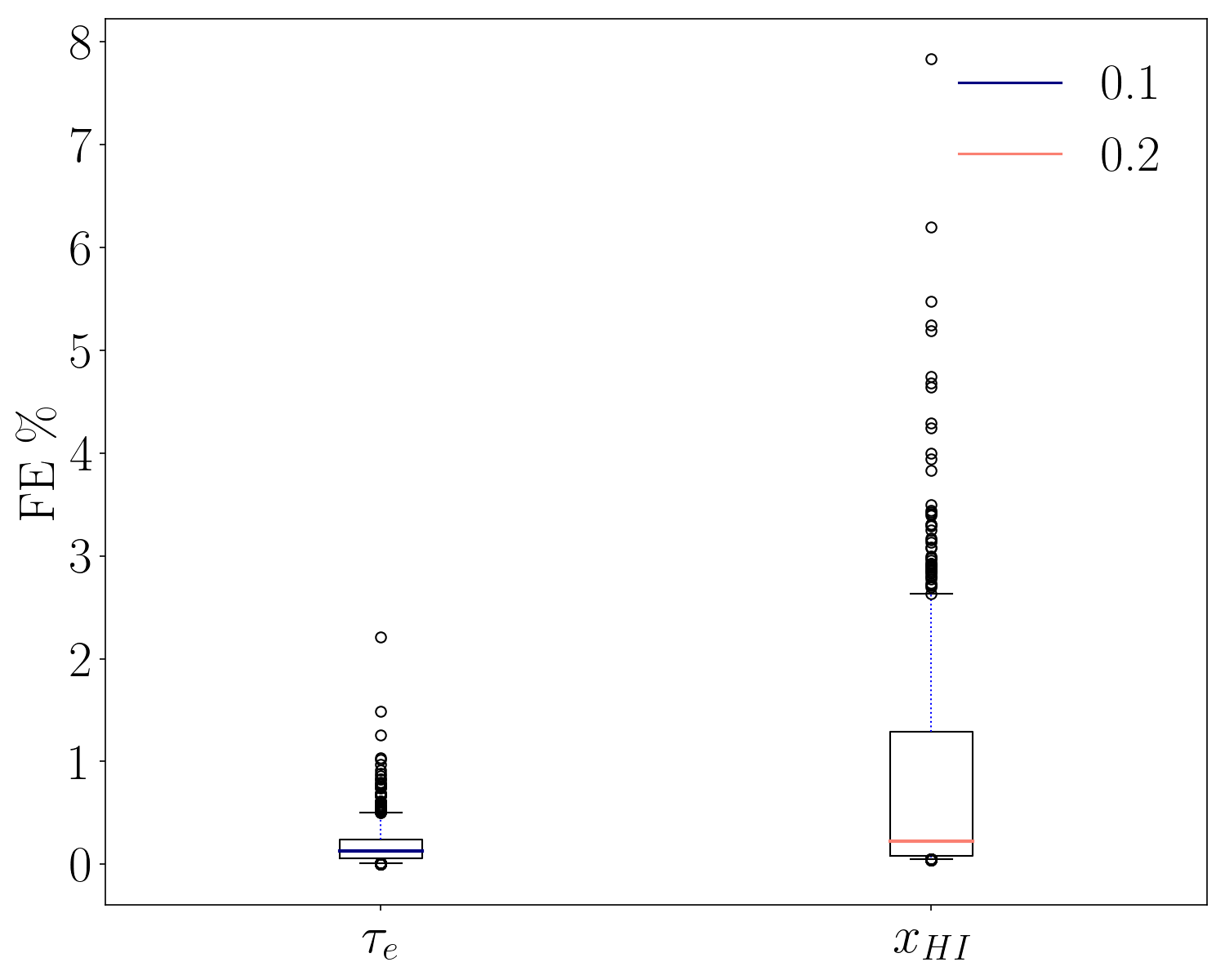}
	\caption{$x_{\rm HI}$ and $\tau_e$ accuracy statistics of the testing set, for $m_{\rm FDM} = 10^{-24}\, {\rm eV}$. The neutral fraction errors are averaged over redshift. The boxes represents all the samples that fall between the 25th and the 75th  FE percentile, and the line in the middle of it is the median. The black circles are outliers that extend beyond the 95th percentile.  The error statistics of the other FDM masses examined here, distributes similarly.}
\label{fig:xHI_tau_testing}
\end{figure}

In Fig.~\ref{fig:UVLF_testing},  the UVLF testing set errors statistics is presented. Our results are similar to the ones achieved in Ref~\cite{Breitman:2023pcj} (see table 1 there) for the same order of magnitude dataset size. This result is reassuring since the dataset \texttt{21cmEMU}~\cite{Breitman:2023pcj} was trained on, was drawn from the MCMC posteriors of Ref.~\cite{HERA:2022wmy}, and as such, most of the samples are centered in the same region is the parameter space. This makes the training process simpler since most of the training samples look alike. This is not the case in our emulator, since the LH sampler draws samples evenly from all around the parameter space. This runs the risk of making the learning process difficult since the variation between different samples can be significant.

We note that $1\sigma$ uncertainty in the HST datapoints ranges from $\sim$ 20\% to $\sim$ 100\%. This implies that an average emulator error of 2-3\% will not affect the inference at all. Fig.~\ref{fig:xHI_tau_testing} shows the same fractional error distributions for $x_{\rm HI}$ and $\tau_e$. In a similar manner to the UVLFs, we conclude that the average errors for these observables is as least one order of magnitude smaller than the standard deviation values quoted in Section~\ref{section_2}, ensuring that those errors will not bias our inference results.  

\begin{figure}[h!]
	\centering
\includegraphics[width = \columnwidth]{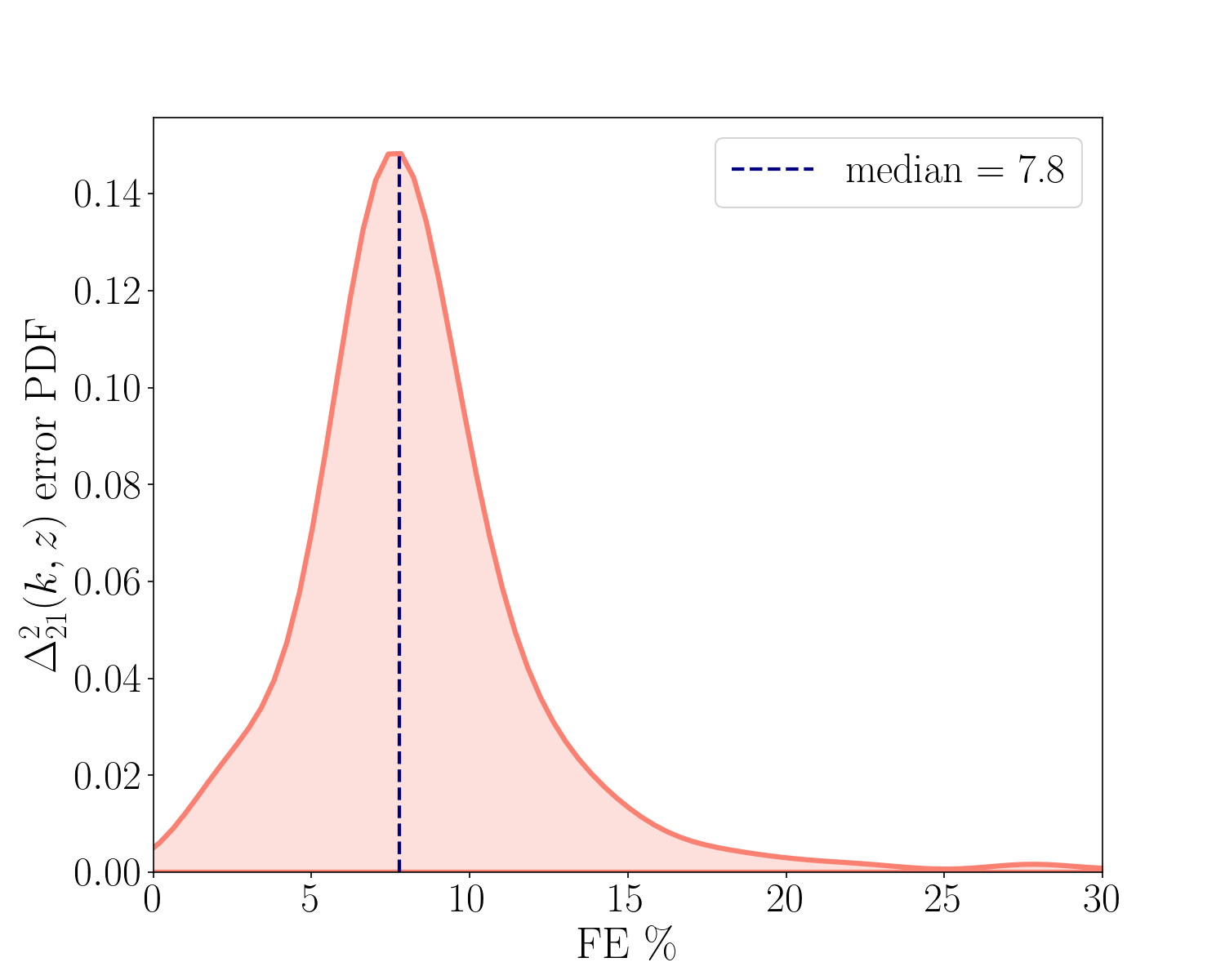}
	\caption{The 21cm power spectrum emulator accuracy statistics of the testing set, averaged over wavenumber and redshift for $m_{\rm FDM} = 10^{-23}\, {\rm eV}$. The error statistics of the other FDM masses we examined here have similar distributions.}
\label{fig:21PS_testing}
\end{figure}

The emulator errors for the 21cm power spectrum are shown in Fig.~\ref{fig:21PS_testing}. Since the current measurements of the power spectrum yield only loose upper bounds, a median error of $\sim 8\%$ will not effect our current study. Future experiments will  probably deliver tighter bounds on the signal, which will require a more accurate emulator. This can be achieved, e.g., using resampling and retraining methods, as described in Ref.~\cite{Lazare:2023jkg}. The 21cm global signal error are not presented here since we do not use them in our inference, but their median error does  not exceed 1\% for each of the FDM masses examined here. Overall, we conclude that our emulator errors are small enough for all the observables in this work, and so they will not compromise the parameter inference results.

\section{Results}\label{section_4}
\subsection{Bayesian inference}\label{section:4_1}

Our main goal in this Section is to perform a MCMC analysis for each of the FDM masses we consider using our emulator pipeline described in Section~\ref{section_3}, to  set an upper bound on the FDM fraction for each mass. To compute the likelihood function we use a Gaussian likelihood for $\tau_e$ and the UVLFs, and a one-sided Gaussian for the neutral hydrogen fraction. For the 21cm power spectrum, we use the likelihood defined in Ref.~\cite{HERA:2021noe}, which treats the current datapoints as upper bounds. 

\begin{table}[h!]
\centering
    
\begin{tabular}{ |c|c|c|}
 \hline
 \multicolumn{3}{|c|}{Priors} \\
 \hline
 Parameter name &Lower bound &Upper bound\\
 \hline 
 $f_{\rm FDM}$ & 0.005 & 0.350 \\
 $H_0$ & 66.82 & 68.50 \\
 
 $\Omega_m$ & 0.3000 & 0.3223 \\
 $10^{10}\ln A_s$ & 3.019 & 3.075\\
 $n_s$ & 0.9589 & 0.9741 \\
 $\log_{10}f_{{\rm esc},10}$&   -3.0  & 0.0\\
 $\alpha_{\rm esc}$& -1.0  & 1.0\\
 
 $\alpha_\ast$ & 0.1  & 1.0\\
 
 $\beta_\ast$& -1.0  & -0.1\\
 
 $\epsilon_\ast^i$& -1.0  & 0\\
 $M_p^i$& 11.0  & 13.0\\
 \hline
 
\end{tabular}

\caption{Prior range summary for all the astrophysical and cosmological parameters used in the MCMC runs.}
\label{tab:prior_table}
\vspace{-0.15in}
\end{table}

The prior distributions for each of the  cosmological and astrophysical parameters, which are summarized in Table~\ref{tab:prior_table}, are all taken to be flat.
The priors ranges for the cosmological parameters, except $f_{\rm FDM}$ are taken to be the $2\sigma$ limits set by the Planck collaboration \cite{Planck:2018vyg}. The MCMC part in our pipeline is implemented using the \texttt{Python} \texttt{Emcee}~\cite{Foreman-Mackey:2012any} sampler, and the corner plots for the posteriors  are generated using the \texttt{corner} \cite{Foreman-Mackey:2016} package.

 \begin{figure}[t!]
	\centering
\includegraphics[width = \columnwidth]{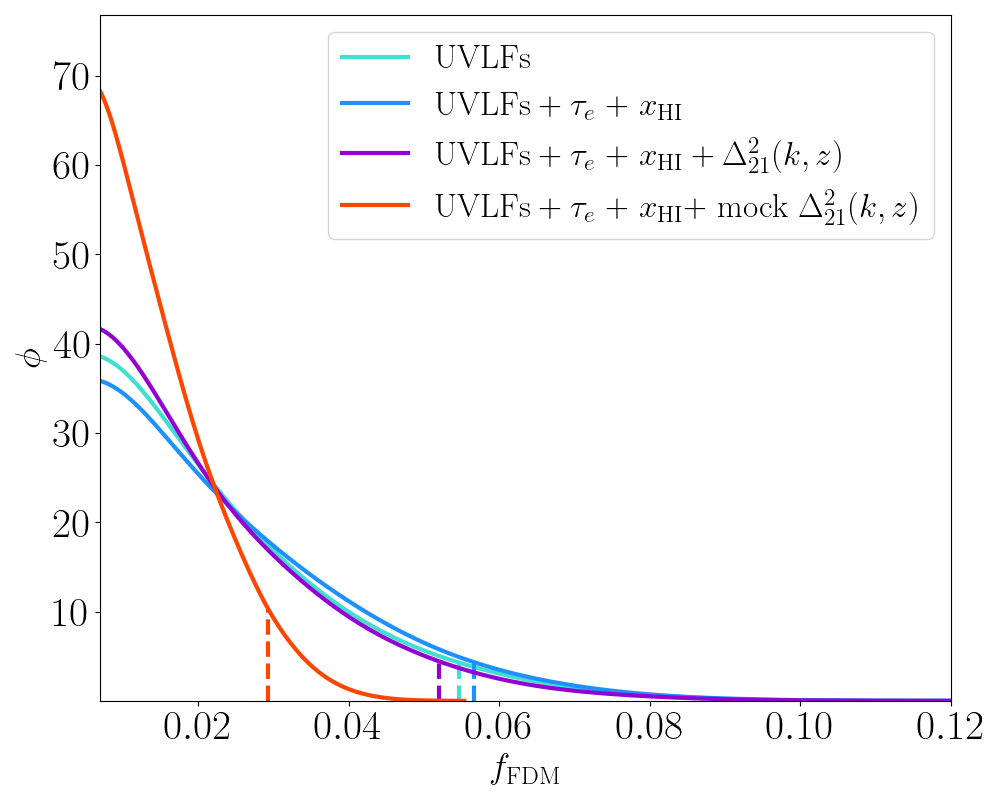}
	\caption{ Posterior distribution for $f_{\rm FDM}$ when using UVLFs separately or together with other observables, for $m_\mathrm{FDM} = 10^{-24} \, \mathrm{eV}$.  The dashed lines represent the 95\% highest posterior density region. The red curve represents the posteriors when using mock data for the 21cm power spectrum as described in section \ref{section 4.2}.}
\label{fig:separate_obs}
\end{figure}

We first examine the constraints imposed by the different observables.
The posteriors for $f_{\rm{FDM}}$ when using the observables separately and together for $m_{\rm{FDM}} = 10^{-24} \, \mathrm{eV}$ are presented in Fig.~\ref{fig:separate_obs}. We note that the UVLFs have most of the constraining power, and adding the other observables does not change the allowed values for $f_{\rm{FDM}}$ for this mass. The same result is obtained for other masses as well. This is reasonable, since UVLFs have many more data points than $\tau_e$ and $x_\mathrm{HI}$ which contribute only one data point each, and the bounds on $\Delta_{21}^2(k,z)$ are loose at the moment, and do not have much constraining power over the other observables, as was shown in Refs.~\cite{HERA:2021noe,Lazare:2023jkg}. We can also see that with future 21cm measurements, as we elaborate on below, the bound on $f_{\rm{FDM}}$ in this mass range is expected to improve.

\begin{figure}[h]
	\centering
\includegraphics[width = \columnwidth]{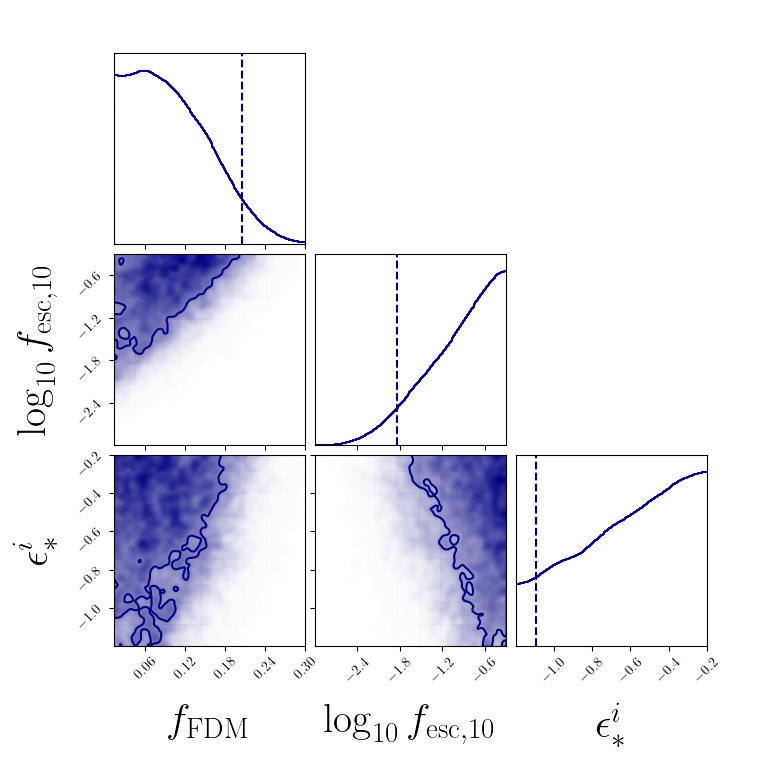}	
\caption{Posteriors  for \{$f_{\rm FDM}$, $f_{\mathrm{esc},10}$, $\epsilon_*^i$\}. The marginalized 2D posteriors show the degeneracy between the  parameters. Dashed lines mark the $95\%$ highest posterior density region. }
\label{fig:reduced_posteriors}
\end{figure}

When using only $\tau_e$ and $x_\mathrm{HI}$ data, the constraints are $~4$ times weaker and the upper bound on the FDM fraction is $~20 \%$, which is an improvement from current constrains at $m_{\rm{FDM}} = 10^{-24} \, \mathrm{eV}$. The reason for this weaker bound, besides the fact that this dataset contains only 2 points, is the degeneracy between the stellar to halo mass ratio $\epsilon_\ast^i$, the escape fraction amplitude $f_{\mathrm{esc}, 10}$ and $f_\mathrm{FDM}$ as shown in Fig.~\ref{fig:reduced_posteriors}. The source of this degeneracy is the fact that reducing $\epsilon_*^i$ and increasing the FDM fraction both result in a smaller star formation rate density (SFRD) (see Ref.~\cite{Munoz:2021psm}), which decreases the number of sources that emit ionizing radiation into the IGM. This can be compensated by a lager ionizing escape fraction, which can be achieved by increasing $f_{\mathrm{esc},10}$.
This is reflected by a wider valid range for the FDM fraction. This degeneracy is broken when accounting for UVLFs, since $\epsilon_\ast^i$ is well constrained by the UVFLs, which leads to constraints on the escape fraction through $\tau_e$ and $x_\mathrm{HI}$. 

\begin{figure}[h!]
	\centering
\includegraphics[width = \columnwidth]{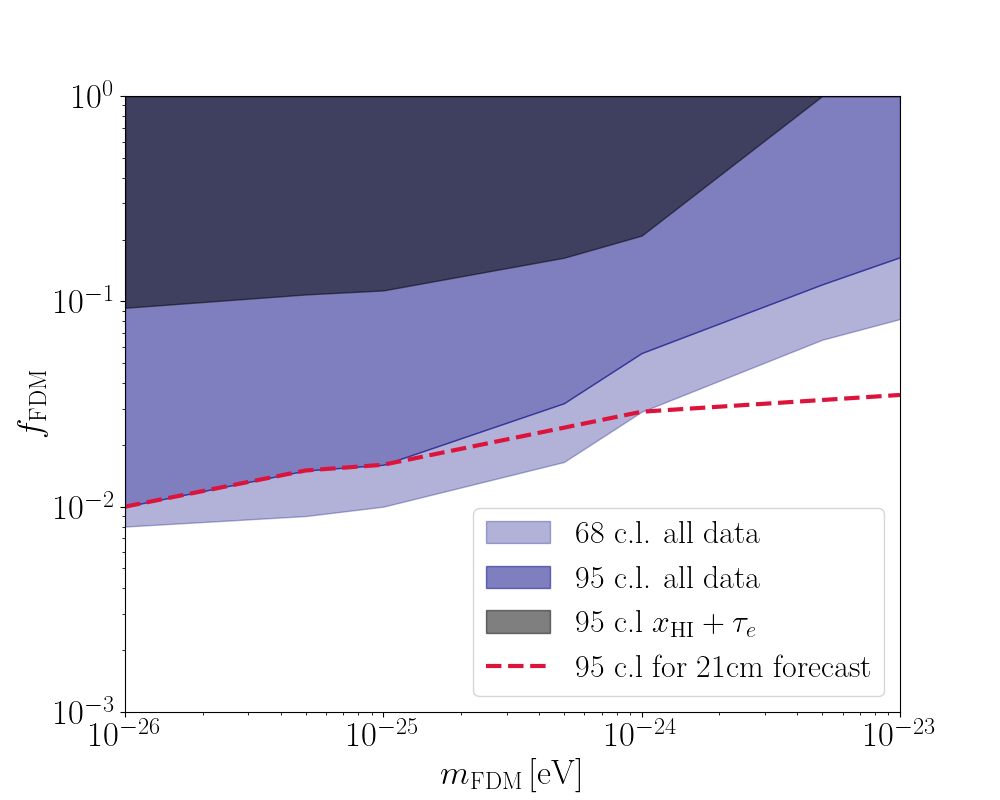}
\vspace{-0.2in}
\caption{Constraints from this work in the two dimensional FDM parameter space. The light and medium blue shaded regions match constraints with 95\% confidence and 68\% confidence levels. The darkest shade corresponds to our constraints using only $x_{\rm HI}$ and $\tau$. The red solid line is our forecast for constraints that can be achieved via a future 21cm power spectrum detection with HERA (see text for details). A future HERA observation can help reduce the upper bound on the FDM fraction for $m_\mathrm{FDM} \!=\! 10^{-23} \, \mathrm{eV}$ from $\sim\!15\%$ to $\sim\!3\%$.}
\label{fig:frac_mass_space_constraints}
\end{figure}

Fig.~\ref{fig:frac_mass_space_constraints} shows the constraints achieved using all the observables in the mass-fraction parameter space examined in this work. These results, summarized in Table~\ref{Tab:mass_bounds}, are a significant improvement over  current bounds in this mass range (see comparison in Fig.~\ref{fig:all_constraints}). The main driver of these new limits are the UVLFs at redshifts $z  = 4 - 10$. 

\begin{table}[h]
\centering
   \begin{tabular}{|| c | c | c ||}
    \hline
    \hline
    $m_{\rm FDM} \, {\rm [eV]}$ & 68\% c.l.  &  95\% c.l. \\
    \hline

$10^{-23}$  &  0.082  &  0.164  \\
$5 \cdot 10^{-24}$  &  0.070  &  0.141  \\
$10^{-24}$  &  0.0288  &  0.056  \\
$5 \cdot 10^{-25}$  &  0.016  &  0.030  \\
$10^{-25}$  &  0.010  &  0.016  \\
$5 \cdot 10^{-26}$  &  0.009  &  0.015  \\
$10^{-26}$  &  0.008 &  0.010  \\

    \hline
    \hline
    \end{tabular}
    \caption{New constraints on FDM for different mass values with 68\% and 95\% confidence level.}
    \label{Tab:mass_bounds}
\end{table}   

As mentioned above, although the current 21cm power spectrum upper bounds have almost no influence on  the derived constrains, results from ongoing and future experiments aiming to detect it, can improve the limits we presented here. In the following we conduct a forecast for the limits that can be achieved using those experiments.

\subsection{Future bounds using a HERA forecast}\label{section 4.2}

In the following we forecast limits on the FDM fraction that could be achieved using a future HERA detection of the 21cm power spectra. We should mention that additional observations beyond those used to derive the upper bounds  above were already conducted in HERA sixth season (2022-2023)~\cite{Breitman:2023pcj}, but are not available publicly so far. 
This observation spans  a wider frequency range, and expands the coverage to cosmic dawn redshifts. 
Future experiments will use the full HERA survey design, which has 330 antennas that will observe the sky for $\sim$ 540 nights.
Here we generate a mock observation for HERA, using our \texttt{21cmFirstCLASS} simulation, with an initial parameter set that is taken to be the median of the MCMC posteriors for the run shown in Section~\ref{section:4_1}.

In order to generate noise for our mock observation, we use \texttt{21cmSense}~\cite{Murray:2024the}, a \texttt{Python} open source repository that forecasts the noise given the antennae configuration and the signal, for the HERA experiment. 

 With the goal of demonstrating the power of 21cm detections to constrain FDM in mind, we choose to use only 3 wavenumber bins ($0.16 \, \mathrm{Mpc}^{-1}$, $0.32 \, \mathrm{Mpc}^{-1}$, $0.48 \, \mathrm{Mpc}^{-1}$) as measured signals, and to adopt conservative 
 assumptions for the experimental configuration and observation time. We assume that the experiment has only 216 functioning antennae that observed the sky for 100 nights, 6 hours per night. The noise for such a configuration is expected to be smaller than in HERA's sixth season, but larger than full HERA.

 For this pipeline, we use the $\tau_e$, $x_{\mathrm{HI}}$ and UVLFs likelihoods as was done so far, but we alter the 21cm power spectra treatment. The likelihood for the mock data no longer treats the 21cm measurements as upper bounds, and is now defined as a Gaussian likelihood. Our emulator predictions can contain a few percentage error, as noted in Fig.~\ref{fig:21PS_testing}, that can have some impact on the inferred constrains. But since this part of our work is merely a forecast, we ignore these errors, and leave the goal of generating a more accurate emulator to future work.

 Fig.~\ref{fig:frac_mass_space_constraints} shows the forecast constrains on the FDM fraction for all the masses when using the 21cm power spectrum likelihood (red curve). This result implies that future 21cm observations have the ability to further constrain the FDM parameter space, beyond what we achieved with UVLFs, $x_{\mathrm{HI}}$ and $\tau_e$, at least for respectively high FDM mass. The constrains forecast for a detected 21cm power spectrum do not feature the same FDM mass dependence as the other observables. The reason that bounds on the FDM fraction do not scale strongly with mass for the 21cm power spectrum, is that the halos that govern the behavior 21cm signal are smaller than the ones that dominate the UVLFs used here. These halos are already affected by FDM with particle masses as low as $\sim \, 10^{-23} \, \mathrm{eV}$. This is not the case for UVLFs, as the halos that control the bright end become more  sensitive to lighter axions.
 This behavior of the fraction limits was already predicted in the study of Ref.~\cite{Flitter:2022pzf}, based on a simple Fisher analysis. Ref \cite{Giri:2022nxq} conducted a similar forecast using a halo-model implementation of the 21cm signal, for the Square Kilometer Array (SKA) telescope, and predicted constraints that match our results.

\section{Conclusions}\label{section_5}

This work presents a machine learning based pipeline for emulating the outputs of \texttt{21cmFirstCLASS}, that can account a presence of ultra light axions. This pipeline includes a modified version of the SFR, based on the prescription introduced in \texttt{GALLUMI} \cite{Sabti:2021xvh, Sabti:2023xwo}. This emulator achieves accurate results for the UVLFs, the optical depth to reionization and the neutral hydrogen fraction, and its errors are at least an order of magnitude smaller than the standard deviations of the measurements. The error for the 21cm power spectra are $\sim 10\%$ which does not affect the inference when using the currently available measurements. Future and on-going experiments will have the ability to reduce the noise and deliver tighter bounds on the power spectra, and so may require a more accurate emulator. This can be achieved using retraining methods, as was demonstrated in Ref.~\cite{Lazare:2023jkg} and Ref.~\cite{Nygaard:2022wri}. 

We used our emulator together with measurements for UVLFs~\cite{Bouwens:2021abc, Gillet:2019fjd}, the CMB optical depth~\cite{Planck:2018vyg}, the neutral hydrogen  fraction~\cite{McGreer:2014qwa} and the 21cm power spectrum~\cite{HERA:2022wmy} and performed an MCMC inference with the objective of constraining the FDM fraction for a given axion mass. Figs.~\ref{fig:LF_z=6_M=-23}--\ref{fig:T21_varing_f_FDM} present the impact of FDM on all the observables used here and demonstrate their power to limit the FDM fraction. The result of this inference is the strongest bounds to date on axion DM in the FDM mass window, limiting its fraction to less than 16\% for $m_{\mathrm{FDM}} = 10^{-23} \, \mathrm{eV}$, and  less than 1\% for $m_{\mathrm{FDM}} = 10^{-26} \, \mathrm{eV}$ with 95\% confidence level, as shown in Table~\ref{Tab:mass_bounds}. This bound shrinks  the available regions in the FDM parameter space and help direct the study of this model to its plausible areas. 
Finally, we conduct a forecast for the bounds that can be achieved with a future detection from 21cm power spectra experiments such as HERA. We find that such observations can reduce the bound on $f_{\mathrm{FDM}}$ by a noticeable amount, for relatively heavier axions, while the lower mass range would be still dominated by the UVLFs. 

\begin{figure}[h!]
	\centering
\includegraphics[width = 0.95\columnwidth]{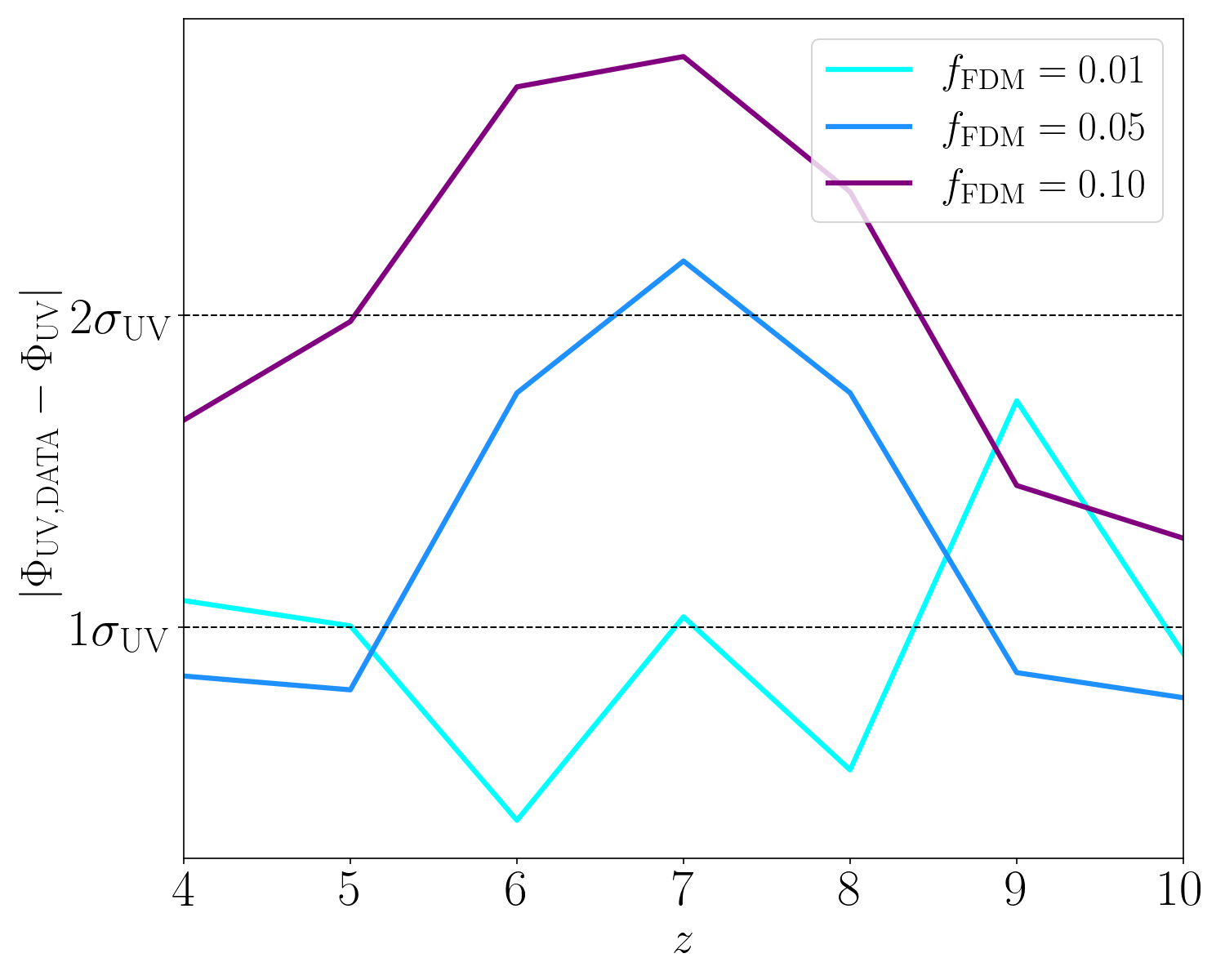}
\vspace{-0.15in}
\caption{The discrepancy between the data and the best fit model with varying $f_{\mathrm{FDM}}$, per $z$, averaged over the magnitude bins, for $m_{\mathrm{FDM}} = 10^{-24} \, \mathrm{eV}$. The difference is presented in units of the standard deviation of the HST measurements. }
\label{fig:errors_per_redshift}
\end{figure}

During the final steps of this work, we encountered Ref.~\cite{Winch:2024mrt} that reached similar conclusions regarding constrains on the FDM parameter space. The main observables used in Ref.~\cite{Winch:2024mrt} are the CMB  temperature power spectrum and the HST UV luminosity functions. Here we do not use the CMB temperature power spectrum likelihood, but the prior range of the cosmological parameters is taken to be the 2$\sigma$ limits from Planck~\cite{Planck:2018vyg}, which has similar impact. In both inferences the main driver of the constrains are the UVLFs. A significant difference between the studies is that we use an emulator instead of the direct approach used in Ref.~\cite{Winch:2024mrt}. The emulator pipeline can
introduce small accuracy errors to the inference, but those are shadowed by the advantages of this method. First, the ANN-based MCMC is significantly accelerated with respect to a computation-based one. This makes our pipeline robust, and drastically simplifies accounting for different, more complicated cosmological models. Furthermore, it enables to take into account in greater detail the observables affected by astrophysics, such as the 21cm power spectrum, the optical depth to reionization and the neutral hydrogen fraction. This is important since those are impacted by the SFR, which is also constrained by the UVLFs. It is crucial to check that the SFR model and parameters that fit the HST UVLFs, do not contradict the other observations, and this can be done easily using an emulator-based MCMC. 
Our work also forecasts bounds that can be derived using future detections of the 21cm signal, e.g.\ by HERA. This is possible thanks to the emulator-based approach, since a direct computation of the 21cm power spectrum can take up to $\mathcal{O}(1 \, \rm{hour} )$, which leads to weeks-to-months--long runtime for MCMC analyses. 
Ref.~\cite{Winch:2024mrt} also examined the impact of UVLFs from JWST. It showed that JWST detections are not yet significant enough to overcome the constraining power of HST. Another difference between our work and Ref.~\cite{Winch:2024mrt}, is that we chose to set a constant value for some of the astrophysical parameters varied in Ref.~\cite{Winch:2024mrt}. The reason for that is that training an emulator on a high-dimensional parameter space requires a large amount of training samples, and each added  parameter contributes exponentially to the number of samples needed for an adequate accuracy. The parameters we decided to keep constant (see Section~\ref{section_3}) have the smallest impact on the luminosity function form, and their influence can be partially compensated by other parameters. Nevertheless this can source some of the differences between our results and that of Ref.~\cite{Winch:2024mrt} (see Appendix~\ref{appendix_A}).

Overall, our results and those published in Ref.~\cite{Winch:2024mrt} agree that FDM is strongly constrained by the HST UVLFs, although there is some difference in the allowed fraction of FDM for each axion mass. Fig.~\ref{fig:errors_per_redshift} shows the discrepancy between the data and  our best fit model, with different values of FDM fraction averaged over $M_{\mathrm{UV}}$, for $m_{\mathrm{FDM}} = 10^{-24} \, \mathrm{eV}$. While considering this plot, one must take into account that at redshifts $z = 9,10$, there are much fewer datapoints than in the later redshifts. With this in mind, one can intuit why our inference prohibits $f_{\mathrm{FDM}} \gtrsim 0.05$ for $m_{\mathrm{FDM}} = 10^{-24} \, \mathrm{eV}$. While for $z = 4, 5$, $f_{\mathrm{FDM}}=0.05$ is slightly more likely, for $z = 6, 7, 8$ the difference between the models fit to the data exceeds $1\sigma$. This will result in a strong favoring of smaller FDM fractions, which provides a visual justification for the results presented in Table~\ref{Tab:mass_bounds}, and further validates our pipeline and conclusions. Based on our findings, it seems  safe to conclude that FDM cannot make up more than $\sim\!10\%$ of  DM for any axion mass smaller than $10^{-23} \, \mathrm{eV}$.

\begin{figure}[hb!]
	\centering
\includegraphics[width = 0.975\columnwidth]{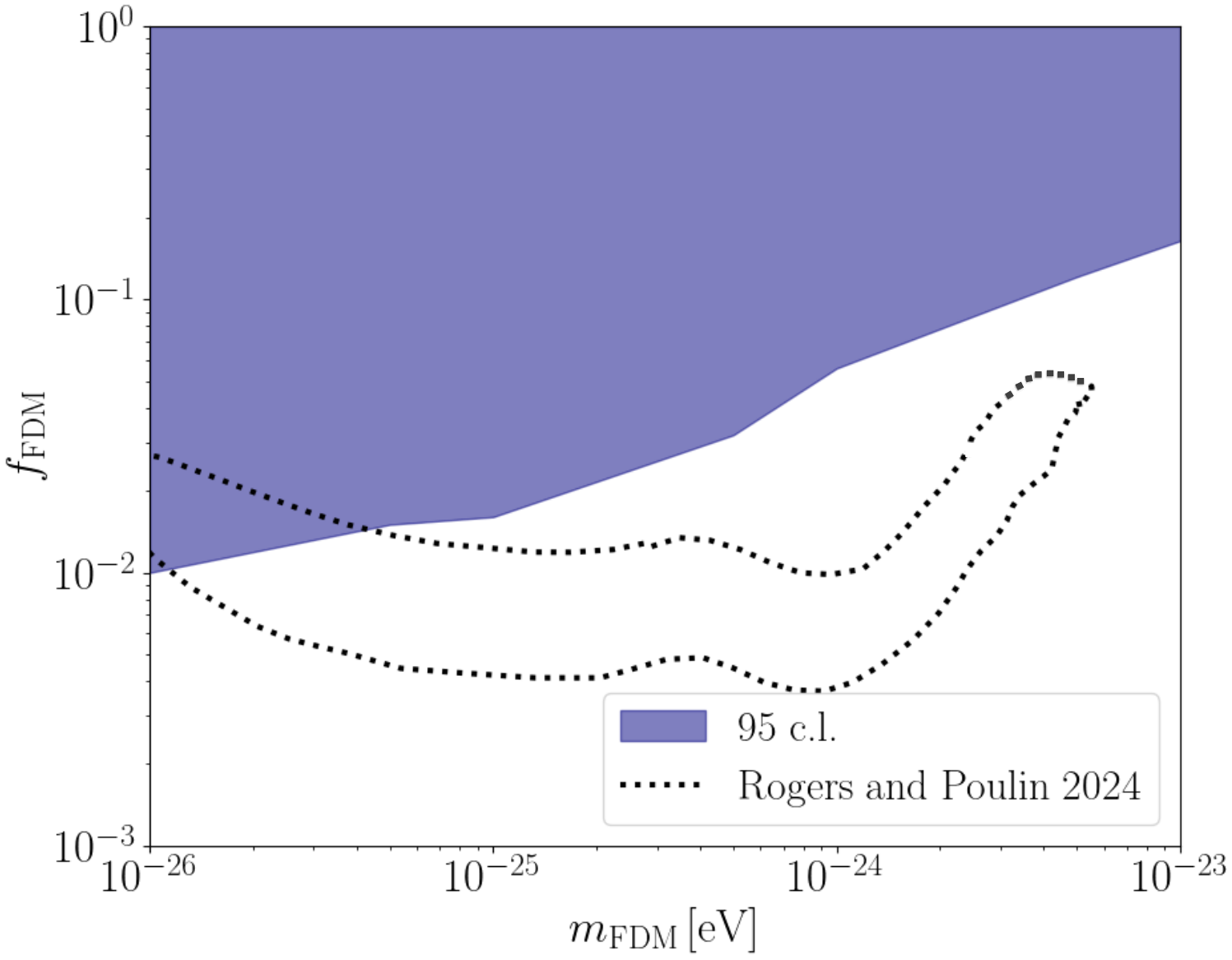}
\vspace{-0.15in}
\caption{95\% c.l.\ constraints from this work alongside the highest preference region for axion FDM reported in Ref.~\cite{Rogers:2023upm}, which used Planck CMB and eBOSS Ly$\alpha$-forest data. Our constraints start to push into this preferred region.}
\label{fig:RogersPoulin}
\end{figure}

Lastly, in Fig.~\ref{fig:RogersPoulin} we plot our final constraints alongside the preferred region found in a  joint analysis of CMB and Ly$\alpha$-forest measurements using data from the Extended Baryon Oscillation Spectroscopic Survey
(eBOSS~\cite{eBOSS:2020yzd}), which found that a non-zero fraction of axion FDM in the $10^{-26}\,\mathrm{eV}\!\lesssim\! m_\mathrm{FDM}\!\lesssim\!10^{-23}\,\mathrm{eV}$ mass window can ameliorate the S8 tension (see Ref.~\cite{Rogers:2023upm}). Most of this region remains allowed by our limits. Future UVLF  and 21cm data should allow to fully test this hypothesis.

\begin{acknowledgements}

The authors wish to thank Daniela Breitman and Sarah Libanore for useful discussions. JF is supported by a Negev PhD Fellowship from Ben-Gurion University. EDK acknowledges joint support from the U.S.-Israel Bi-national Science Foundation (BSF, grant No.\ 2022743) and the U.S.\ National Science Foundation (NSF, grant No.\ 2307354), and support from the ISF-NSFC joint research program (grant No.\ 3156/23). 

\end{acknowledgements}

\appendix
\section{MCMC posteriors}\label{appendix_A}

\begin{figure*}
    \centering
\includegraphics[width =\textwidth]{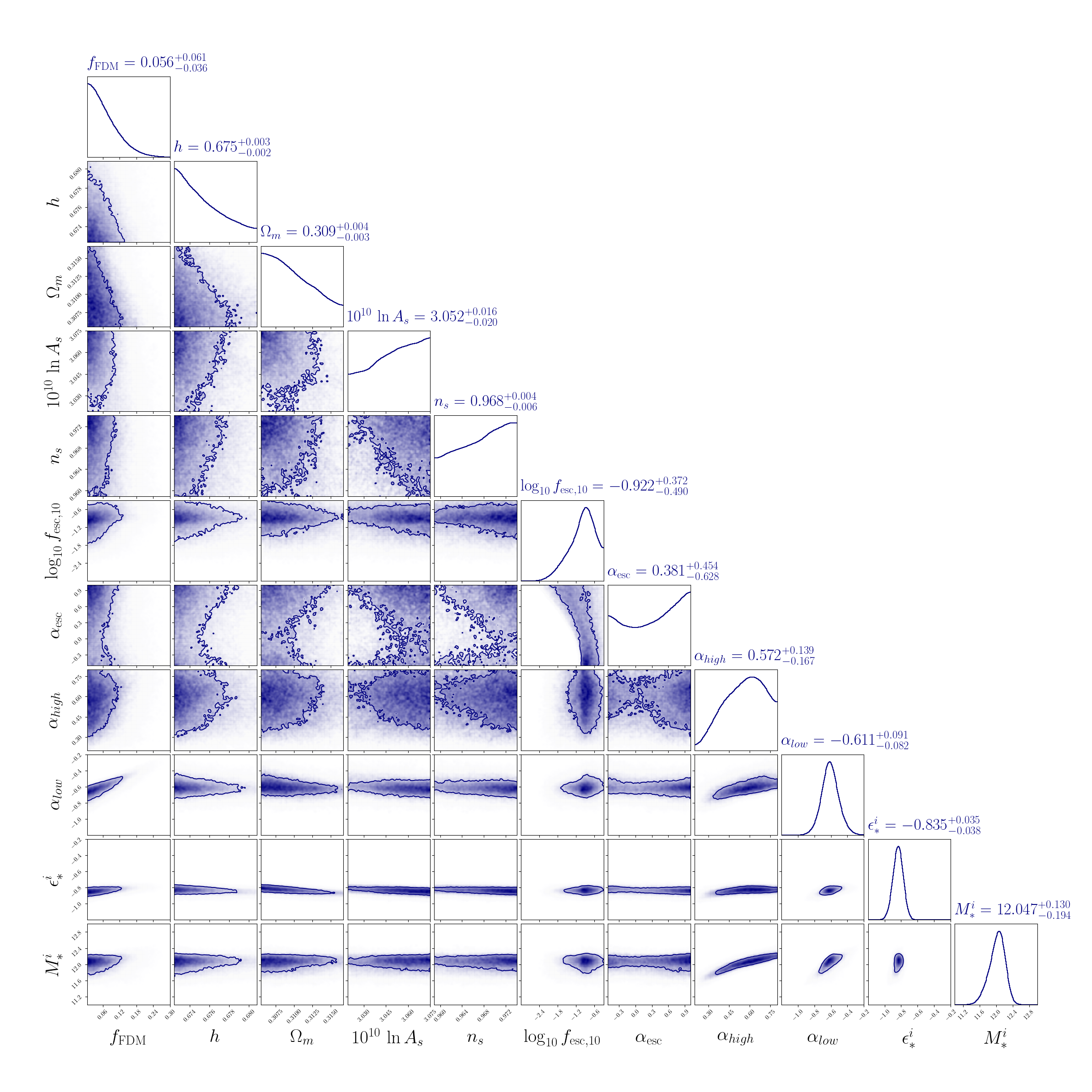}\\
\vspace{-0.1in}
\caption{Posterior distribution for all the model initial parameters, both cosmological and astrophysical, for $m_{\mathrm{FDM}} = 10^{-23} \mathrm{eV}.$}
\label{fig:complete_posterior}
\end{figure*}

Here for completion we present the posterior distribution of the inference described in Section~\ref{section:4_1}, for all the astrophysical and cosmological parameters. Fig.~\ref{fig:complete_posterior} shows the posterior distribution for $m_\mathrm{FDM}=10^{-23} \, \mathrm{eV}$ with all the observables together. We note that most of the astrophysical parameters are well constrained by the UVLFs and by $\tau_e$, $x_{\mathrm{HI}}$.

\begin{figure}[h]
	\centering
\includegraphics[width = \columnwidth]{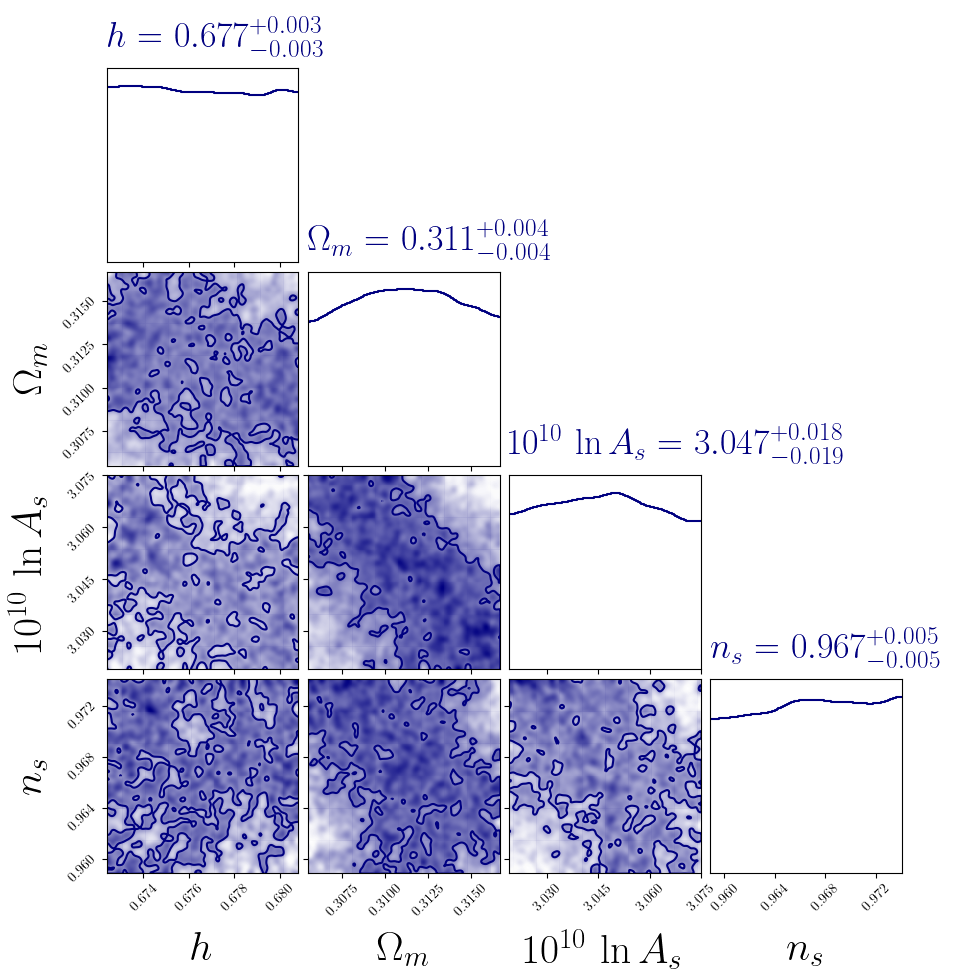}
\caption{Posterior distribution of the cosmological parameters when not accounting for the Alcock-Paczynski effect, for $m_{\mathrm{FDM}} = 10^{-23} \, \mathrm{eV}$. All the posterior features of these parameters vanish when this effect is not considered. The posteriors for all the other parameters remain the same.} 
\label{fig:AP_effect}
\end{figure}

The cosmological parameters exhibit some non-trivial behavior, as smaller or larger values are slightly preferred depending on the parameter.
As shown in Fig.~\ref{fig:AP_effect}. This behavior does not occur when the Alcock-Paczynski effect is not taken into account. The Hubble parameter $h$ and dark matter energy density  $\Omega_m$  are free parameters that impact the data correction this effect generates (see Ref.~\cite{Sabti:2021xvh}), so naturally they experience some constraining power when it is considered. The amplitude $A_s$ and spectral index $n_s$ of primordial fluctuations are degenerate with $h$ and $\Omega_m$ through the luminosity function, so a weak preference of some values of the former, results in a similar effect on the latter. 
We do not treat these as new constraints, as they are not prominent enough to rule out certain values. In addition, the high likelihood combinations must be considered against the CMB power spectrum likelihood, as some of them might be ruled out. This fact does not impair the derived constrains on FDM, as exploring a wider parameter space in principle can only lower the bounds.

\end{document}